\documentclass[journal]{IEEEtran}
\usepackage{commath,graphicx,afterpage,subfigure,amsmath,amsfonts,amssymb,algorithm,algorithmic}
\usepackage{mathrsfs,amsthm,cite,multirow,bm,multicol,booktabs,threeparttable,extpfeil}
\usepackage{dblfloatfix}
\usepackage{cuted}
\usepackage{capt-of}

\newtheorem{definition}{Definition}
\newtheorem{assumption}{Assumption}
\newtheorem{lemma}{Lemma}
\newtheorem{theorem}{Theorem}
\makeatletter
\let\SAVEDmaketitle\maketitle
\let\SAVEDatmaketitle\@maketitle
\makeatother
\begin{document}

\title{Backpropagation-Free Test-Time Adaptation for Lightweight EEG-Based Brain-Computer Interfaces}

\author{Siyang~Li, Jiayi~Ouyang, Zhenyao~Cui, Ziwei~Wang, Tianwang~Jia, Feng~Wan, and Dongrui~Wu, \IEEEmembership{Fellow,~IEEE}
\thanks{This research was supported by Guangdong Major Project of Basic Research (2025B0303000003), National Natural Science Foundation of China (62525305 and 625B2077), and Major Science and Technology Project of Industry-university-research Cooperation for Colleges and Universities in Hebei Province in Shijiazhuang City (2511303107A).}
\thanks{S.~Li, J.~Ouyang, Z.~Cui, Z.~Wang, T.~Jia, and D.~Wu are with the Hubei Key Laboratory of Brain-inspired Intelligent Systems, School of Artificial Intelligence and Automation, Huazhong University of Science and Technology, Wuhan 430074, China. They are also with the Shenzhen Huazhong University of Science and Technology Research Institute, Shenzhen, China.}
\thanks{F.~Wan is with the Department of Electrical and Computer Engineering, Faculty of Science and Technology, University of Macau, Macau 999078, China, and also with the Centre for Cognitive and Brain Sciences, Institute of Collaborative Innovation, University of Macau, Macau 999078, China.}
\thanks{Siyang Li, Jiayi Ouyang, and Zhenyao Cui contributed equally to this work.}
\thanks{Corresponding Author: Dongrui Wu (drwu09@gmail.com).}}
\markboth{IEEE JOURNAL OF BIOMEDICAL AND HEALTH INFORMATICS, 2026}
{LI \MakeLowercase{\textit{et al.}}: BACKPROPAGATION-FREE TEST-TIME ADAPTATION FOR LIGHTWEIGHT EEG-BASED BCIS}
\maketitle

\begin{abstract}
Electroencephalogram (EEG)-based brain-computer interfaces (BCIs) face significant deployment challenges due to inter-subject variability, signal non-stationarity, and computational constraints. While test-time adaptation (TTA) mitigates distribution shifts under online data streams without per-use calibration sessions, existing TTA approaches heavily rely on explicitly defined loss objectives that require backpropagation for updating model parameters, which incurs computational overhead, privacy risks, and sensitivity to noisy data streams. This paper proposes Backpropagation-Free Transformations (BFT), a TTA approach for EEG decoding that avoids these issues. BFT applies multiple sample-wise transformations, based on knowledge-guided augmentations or structured feature masking, to each test trial, producing multiple predictions for a single test sample using only forward passes. A learning-to-rank module, trained on source data, estimates the reliability of each transformed prediction, so that a weighted aggregation suppresses prediction uncertainty during online inference, with theoretical justification. Extensive experiments on five EEG datasets, covering motor imagery classification and driver drowsiness regression, demonstrate the effectiveness, versatility, robustness, and efficiency of BFT. This research enables lightweight plug-and-play BCIs on resource-constrained devices, broadening the real-world deployment of EEG-based BCIs.
\end{abstract}

\begin{IEEEkeywords}
Brain-computer interface, domain adaptation, electroencephalogram, test-time adaptation, transfer learning
\end{IEEEkeywords}

\section{Introduction}

\IEEEPARstart{B}{rain-computer} interfaces (BCIs) translate neural activity into control commands that enable direct interaction between users and external systems, and are increasingly evolving toward closed-loop platforms for neurorehabilitation and cognitive enhancement~\cite{Gao2025}. Non-invasive BCIs, which typically rely on electroencephalography (EEG) sensors, remain the most accessible. In active BCIs, users perform motor imagery (MI) by mentally rehearsing limb movements, and the resulting EEG is decoded in real time into control commands for devices such as prosthetics, exoskeletons, or computer cursors~\cite{Wu2022}. Beyond active control, passive BCIs monitor cognitive states such as emotion~\cite{Wu2023} and driver reaction-time or drowsiness~\cite{Fu2024,Parekkattil2025}, where real-time decoding offers substantial benefits for safety-critical tasks such as driving.

Despite being surgery-free and relatively low cost, EEG suffers from high inter-subject variability and nonstationarity: responses vary significantly between users, and even across sessions of the same user, due to fluctuations in mental state, concentration, or electrode contact quality~\cite{Vidaurre2006}. Consequently, most EEG decoders require lengthy calibration sessions before each use, limiting their practicality in real-world deployments.

Transfer learning (TL)~\cite{Zhuang2021} can reduce or eliminate calibration by leveraging auxiliary data from additional subjects. While classic TL assumes an offline transductive setting, test-time adaptation (TTA)~\cite{Liang2025, Wang2024TTA} supports a more practical online setting in which the model adapts sequentially to streaming test data, making it well-suited to real-time, calibration-free, plug-and-play BCIs.

\begin{figure}[htpb] \centering
\includegraphics[width=1.02\linewidth,clip]{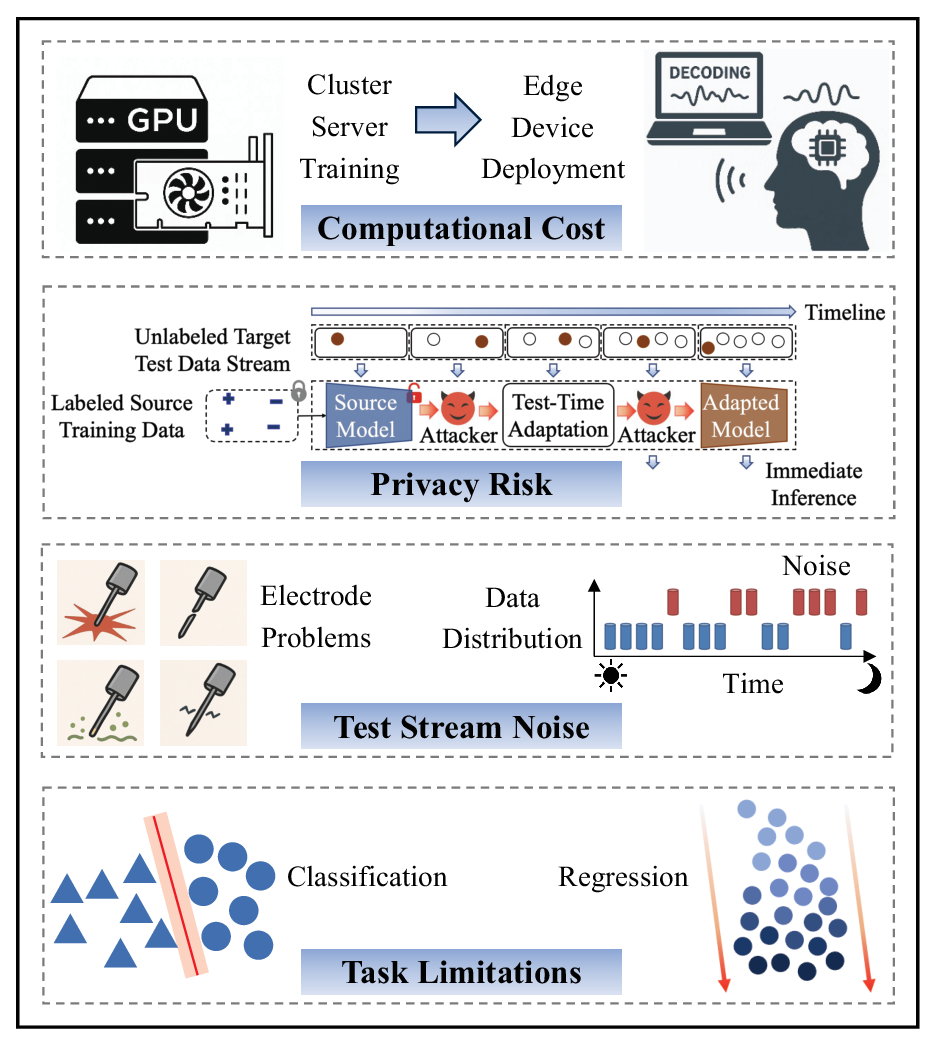}
\caption{Key issues in deploying TTA algorithms for BCI decoding.} \label{fig:issues}
\end{figure}

Nonetheless, the mechanism that makes recent TTA approaches accurate also keeps them off real devices. A majority of TTA approaches adapt by backpropagating a loss objective to update model parameters~\cite{Li2024T-TIME}, and this single choice raises four coupled obstacles, illustrated in Fig.~\ref{fig:issues}. First, gradient updates are costly and often infeasible on the low-power, memory-limited processors of edge BCIs, especially once the model is quantized~\cite{Jacob2018}. Second, updating weights requires white-box access to the model, which exposes it and precludes black-box, privacy-preserving deployment. Third, the update follows whatever test samples arrive, so frequent EEG artifacts can steer it into negative transfer. Fourth, its objectives, such as entropy or class-probability sharpening, are defined for classification and do not extend to regression tasks. These obstacles motivate a TTA approach that is backpropagation-free, privacy-preserving, noise-robust, and task-agnostic.

This paper introduces Backpropagation-Free Transformations (BFT), a TTA approach for online EEG-based BCI deployment under constrained computational resources. BFT applies multiple sample-wise transformations to each test trial and aggregates their predictions to implicitly reduce inference uncertainty, using only forward passes and without updating any model parameter. To exploit the transformations selectively, a learning-to-rank module trained on source data estimates the reliability of each transformed prediction, so that more reliable transformations contribute more to the aggregated output. The same reliability-aware aggregation serves both classification and regression. Extensive experiments on three motor-imagery classification and two driver-drowsiness regression EEG datasets demonstrate that BFT is more practical than existing TTA approaches for lightweight, plug-and-play BCIs.

Our main contributions are summarized as follows:
\begin{enumerate}
\item Proposal of BFT, a lightweight TTA approach that is backpropagation-free, privacy-preserving, noise-robust, and task-agnostic, adapting each prediction using only forward passes of a fixed source model.
\item A unified reliability-aware aggregation that serves both classification and regression through a source-trained learning-to-rank module, with a variance-based theoretical analysis of when it reduces prediction uncertainty.
\item Comprehensive experiments under real-time online streams, structured test-time artifacts, multiple paradigms, and post-training quantization, verifying the effectiveness, versatility, robustness, and efficiency of BFT, and demonstrating that high-performance online decoding can be deployed in plug-and-play EEG-based BCIs without per-use calibration to broaden real-world applicability.
\end{enumerate}

The remainder of this paper is organized as follows: Section~\ref{sect:relatedwork} introduces related work. Section~\ref{sect:approach} proposes BFT. Section~\ref{sect:experiments} presents experimental results. Finally, Section~\ref{sect:conclusions} draws conclusions and points out future research directions. The variance-based theoretical analysis is provided in the Supplementary Material.

\section{Related Work}\label{sect:relatedwork}

This section reviews TTA approaches. TL approaches for unsupervised domain adaptation (UDA) restricted to the offline transductive setting have been comprehensively discussed by Li \emph{et al.}~\cite{Li2024T-TIME} and are thus omitted here. Such approaches are still compared in the experiments to demonstrate offline TL capabilities.

\subsection{Transfer Learning}
Conventional machine learning assumes that training and test sets are independently and identically distributed (i.i.d.), drawn from the same underlying distribution. TL~\cite{Zhuang2021} relaxes this assumption by leveraging knowledge from the source domain to improve performance on the target domain under distribution shift. This field of study is also known as domain adaptation~\cite{Zhuang2021} or concept drift~\cite{Lu2019}.

TL typically addresses three types of distribution shift:
\begin{enumerate}
\item \textbf{Marginal Distribution Shift:} $P_s(\mathbf{x}) \neq P_t(\mathbf{x})$, i.e., changes in the input distribution.
\item \textbf{Conditional Distribution Shift:} $P_s(y|\mathbf{x}) \neq P_t(y|\mathbf{x})$, i.e., changes in the prediction function.
\item \textbf{Label Distribution Shift:} $P_s(y) \neq P_t(y)$, i.e., changes in class priors.
\end{enumerate}

\subsection{Test-Time Adaptation}
In real-time BCIs, test samples arrive sequentially and require low-latency inferences. As a result, the classic UDA setting is inapplicable. TTA~\cite{Liang2025, Wang2024TTA, Xiao2024} provides a more practical alternative and can be viewed as a constrained form of UDA, characterized by:
\begin{enumerate}
\item No access to source data, and only the pretrained source model is available. This is the key distinction between source-free UDA and vanilla UDA, and it also applies to the TTA setting in general.
\item Access is restricted to a small subset of unlabeled target samples at any given time. Unlike UDA, no abundant unlabeled target data are pre-accessible for estimating the target distribution.
\item Iterative optimization is avoided due to computational constraints, as each inference must be returned almost immediately, and physical devices at deployment may not be high-end.
\end{enumerate}

The representative TTA approaches are summarized and categorized in the following paragraphs:

\textbf{TTA for Mitigating Marginal Distribution Shift.} Batch Normalization test-time adaptation (BN-adapt)~\cite{Schneider2020BNadapt} is the most straightforward approach. Test entropy minimization (Tent)~\cite{Wang2021TENT} also updates the batch normalization layers, but through minimizing the entropy of model predictions on test inputs using backpropagation. For EEG data, Euclidean Alignment (EA)~\cite{He2020EA} normalizes the mean covariance matrices of each domain to the identity matrix, and Li \emph{et al.}~\cite{Li2024T-TIME} showed that EA can be seamlessly applied to TTA, with an online updated target reference matrix.

\textbf{TTA for Mitigating Conditional Distribution Shift.} Target Pseudo-Labels (PL)~\cite{DongHyun2013PL} is the most straightforward approach. Uncertainty minimization can implicitly mitigate conditional distribution shift. Sharpness-aware and reliable entropy minimization (SAR)~\cite{Niu2023SAR} selects samples with smaller entropy losses and jointly minimizes the sharpness of the entropy and the entropy loss for a more reliable adaptation. Test-Time Information Maximization Ensemble (T-TIME)~\cite{Li2024T-TIME} extends the information maximization loss objective, which incorporates an additional uniform regularization of label distribution into classic conditional entropy, to TTA.

\textbf{TTA for Mitigating Label Distribution Shift.} Label shift is a difficult problem, and often has to resort to pseudo-labels for estimating statistics of the target label distribution. Marginal Entropy Minimization with One test point (MEMO)~\cite{Zhang2022MEMO} regularizes the model to produce similar predictions for each transformation through mean entropy minimization. Li \emph{et al.}~\cite{Li2024T-TIME} incorporated label shift into information maximization through online estimation.

\subsection{Test-Time Adaptation Beyond Model Update}\label{sect:ttadeficiency}

Despite recent advances in decoding performance~\cite{Li2024T-TIME}, deploying TTA algorithms in BCIs remains challenging due to several practical constraints, as illustrated in Fig.~\ref{fig:issues}.

\textbf{Computational Cost.} TTA approaches often rely on loss objectives under backpropagation, which is infeasible on low-power BCI devices lacking dedicated GPUs. Alternative parameter-learning rules, including Hamilton-Jacobi-Bellman-equation-based optimal learning for BCI models~\cite{Reddy2020HJB}, can avoid conventional gradient backpropagation during training but still update model parameters. By contrast, the deployment setting considered here permits forward inference only. This issue is exacerbated by model quantization~\cite{Jacob2018}, such as converting EEGNet from 32-bit to 8-bit integers for deployment~\cite{Schneider2020}, which hinders retraining or fine-tuning with backpropagation.

\textbf{Privacy Risk.} Updating model parameters during inference requires access to internal weights, exposing sensitive information. Black-box deployment is much more preferable for preserving model privacy~\cite{Xia2023, Dionysiou2023}.

\textbf{Test Stream Noise.} EEG is highly susceptible to artifacts caused by fatigue, movement, sweat, poor electrode contact, etc. Such noise increases the difficulties of hyperparameter selection, model selection, and the combination of different types of shifts for TTA approaches~\cite{Zhao2023}, which could lead to negative transfer when not appropriately handled~\cite{Zhang2023}.

\textbf{Task Limitations.} Most TTA approaches, and TL approaches more broadly, are designed for classification and rely on predicted class probabilities, which restricts their applicability to regression tasks. Approaches that address conditional or label distribution shifts in regression remain largely unexplored.

These limitations highlight the urgent need for a TTA framework that is backpropagation-free, privacy-preserving, noise-robust, and task-agnostic. While a few recent methods remove the need for backpropagation, they offer limited gains and remain unsuitable for regression.

\section{Backpropagation-Free Transformations \\for Test-Time Adaptation} \label{sect:approach}

This section presents the proposed BFT method, which enables TTA without requiring access to model parameters, gradients, or batched inputs during inference.

\subsection{Problem Formulation}\label{sect:formulation}

Let $\mathcal{D}_{\text{test}} = \{\mathbf{x}_t\}_{t=1}^{n_t}$ denote a streaming test set, where $\mathbf{x}_t$ is a test input. The deployed model is composed of a feature extractor $g(\cdot)$ and a task-specific classifier or regressor $h(\cdot)$, trained on a training set $\mathcal{D}_{\text{train}} = \{(\mathbf{x}_i, y_i)\}_{i=1}^{n_s}$, which is assumed to be unavailable at test time.

Due to distributional shifts between the source and target domains (e.g., across subjects or sessions), model performance may degrade at test time. TTA addresses this degradation by refining predictions during online inference. At each time step $t$, TTA aims to improve the prediction $\hat{y}_t$ using only $\{\mathbf{x}_t, \hat{y}_t, g, h\}$, i.e., the current test input $\mathbf{x}_t$, its initial prediction $\hat{y}_t=h(g(\mathbf{x}_t))$, and the frozen source-trained feature extractor $g$ and task head $h$, without access to the training data, ground-truth labels, or gradients.

\subsection{Test-Time Transformations}\label{sect:TTT}

Uncertainty estimation is widely employed in TL, particularly to implicitly address conditional distribution shifts beyond marginal distribution shifts. Shannon entropy~\cite{Shannon1948}, derived from the softmax output of a classifier, serves as a representative~\cite{Grandvalet2004}: high entropy indicates domain mismatch, while low entropy suggests alignment. However, entropy minimization necessitates backpropagation and is therefore inapplicable in forward-only settings or regression tasks.

To overcome this limitation, we propose test-time transformations that can be considered as structured perturbations. Intuitively, if a model is well-aligned to the target domain, its predictions should remain stable under such perturbations. Thus, the variability of predictions across transformed inputs can be used as a surrogate measure of uncertainty. A more detailed theoretical derivation is offered in Section~S-I of the Supplementary Material.

\begin{figure}[htpb]\centering
\subfigure[]{\includegraphics[width=\linewidth,clip]{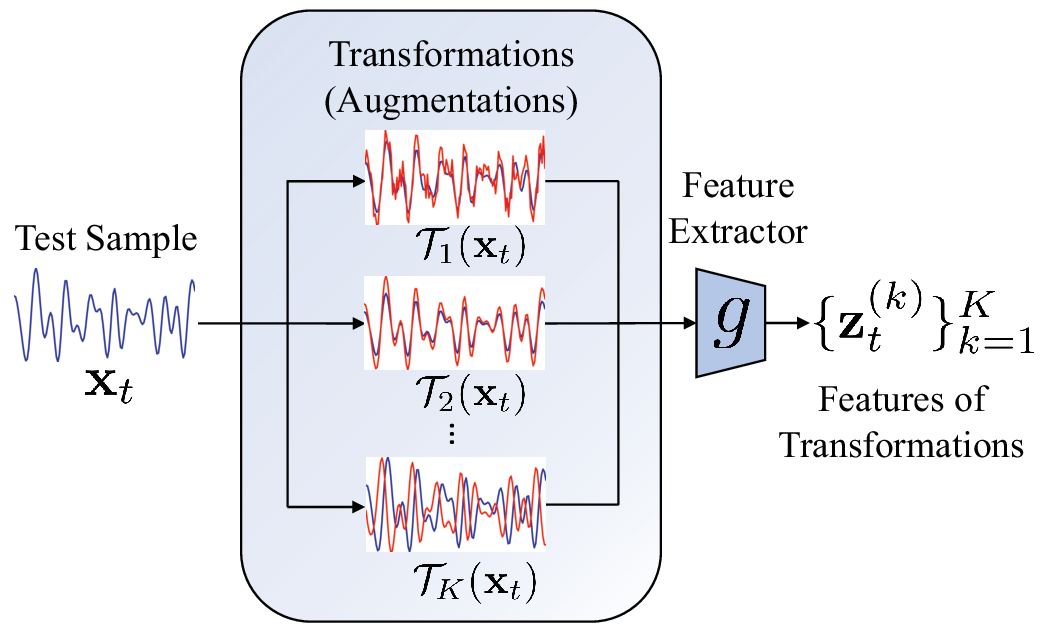}} \hfill
\subfigure[]{\includegraphics[width=.95\linewidth,clip]{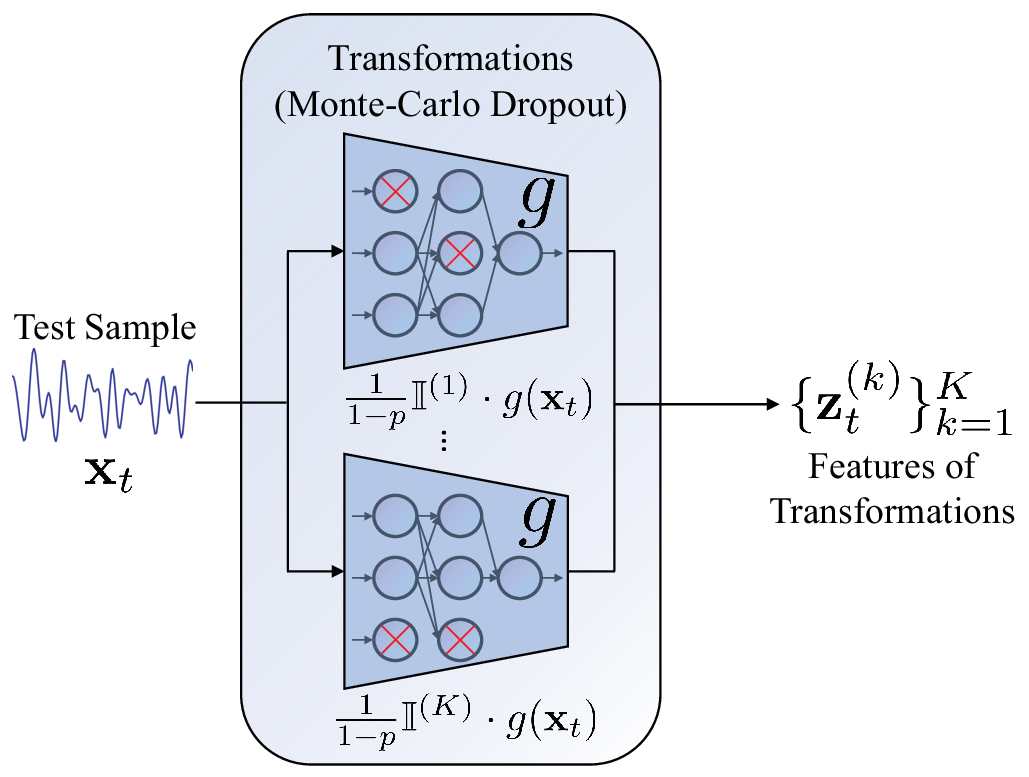}}
\caption{Two types of transformations. (a) BFT-A; and (b) BFT-D.} \label{fig:transformations}
\end{figure}

We consider two types of transformations, with illustrations shown in Fig.~\ref{fig:transformations}.

\textbf{Knowledge-guided Augmentations} (referred to as BFT-A): These transformations are commonly employed for EEG data augmentation~\cite{Freer2020}. Examples include:
\begin{enumerate}
\item \textit{Noise Addition (Noise):} Injects uniform noise into the input signal.
\item \textit{Amplitude Scaling (Scale):} Multiplies the signal by a scalar close to one to slightly adjust its amplitude.
\item \textit{Frequency Shift (Freq):} Uses the Hilbert transform to shift the signal's frequency content.
\item \textit{Sliding Window (Slide):} Generates overlapping temporal segments from each trial using a sliding window.
\end{enumerate}
Such augmentations are a set of $K$ deterministic or stochastic transformations, denoted as $\{\mathcal{T}_k(\cdot)\}_{k=1}^K$, applied to input $\mathbf{x}_t$. The resulting features of each of the transformations are:
\begin{align}
\mathbf{z}_t^{(k)} = g(\mathcal{T}_k(\mathbf{x}_t)). \label{eq:BFT-A}
\end{align}

\textbf{Deterministic Dropout Subnetwork Bank} (referred to as BFT-D): Inspired by Monte-Carlo dropout~\cite{Gal2016}, BFT-D constructs a fixed bank of feature-masked subnetworks. Unlike conventional Monte-Carlo dropout, it does not resample a Bernoulli mask for every trial. Each branch retains a stable identity, allowing the ranking module trained for that branch to be applied consistently at test time.

Assuming that a dropout layer exists after $g(\cdot)$, each branch employs a binary mask $\mathbb{I}^{(k)} \in \{0,1\}^d$ applied to the feature vector $g(\mathbf{x}_t) \in \mathbb{R}^d$. Each mask deterministically drops a fixed subset of features. This fixed mask bank preserves the train-test correspondence of branch $k$, which is required by the branch-specific ranking formulation in the following subsection. Replacing it with independently resampled masks would change that correspondence and therefore define a different method. The mask values are:
\begin{align}
\mathbb{I}^{(k)}_i = 
\begin{cases}
0, & \text{if } i \in \left[ (k{-}1)\frac{d}{K},\ k\frac{d}{K} \right), \\
1, & \text{otherwise},
\end{cases} \label{eq:dropoutmask}
\end{align}
where $\frac{1}{K}$ corresponds to the original training-time dropout rate $p$. The masks therefore generate different, repeatable feature subsets of the same test sample. Features may also be dropped non-consecutively or shared across masks, provided that the mask bank is fixed between ranking-module training and deployment.

The resulting feature of each of the transformations is:
\begin{align}
\mathbf{z}_t^{(k)} = \frac{1}{1 - p} \mathbb{I}^{(k)} \cdot g(\mathbf{x}_t), \label{eq:BFT-D}
\end{align}
where the scaling factor $\frac{1}{1 - p}$ compensates for the reduced activation magnitude, thereby preserving the expected value of the feature vector under the masking, similar to that of training-time dropout.

Note that BFT-A modifies input data, whereas BFT-D alters features. Although both BFT-A and BFT-D require forward passes of multiple samples instead of the original test sample, both are computationally efficient due to batched forward passes under matrix operations. The original test sample's feature may also be included, as the identity transformation.

The representations of the transformations $\{\mathbf{z}_t^{(k)}\}_{k=1}^{K}$ are then forwarded through $h(\cdot)$ to produce multiple predictions for the same test sample $\mathbf{x}_t$.

\subsection{Learning-to-Rank Transformations} \label{sect:rankingmodule}

Not all transformations produce equally reliable predictions. Simple aggregation schemes that assign uniform weights to all transformed outputs fail to account for the varying reliability levels of each transformation. To address this, we propose estimating reliability scores for each transformed input to enable a weighted combination. Inspired by learning-to-rank approaches~\cite{Liu2009}, we further introduce a ranking-based strategy.

Consider a neural network module for ranking that receives feature representations from $g(\cdot)$ and outputs a scalar reliability score in a continuous space, analogous to a regression model. This ranking module, denoted as $r(\cdot)$, can be built on the transformations of training samples. Na\"ively, the reliability scores of these transformations can be simply based on the task losses, obtained using the trained classifier/regressor $h(\cdot)$. However, task losses alone are suboptimal for modeling transformation reliability due to several limitations:
\begin{itemize}
\item No additional information is introduced, as $r(\cdot)$ merely replicates/distills the knowledge embedded in $h(\cdot)$.
\item The loss values are typically close in magnitude since $h(g(\cdot))$ is optimized on the training data, thereby impeding the effective optimization of $r(\cdot)$.
\item Lower task loss values would correspond to higher reliability, which is inversely correlated.
\item Task losses ignore the relative relationship across different transformations of the same instance.
\end{itemize}

\begin{figure}[htpb] \centering
\includegraphics[width=1.05\linewidth,clip]{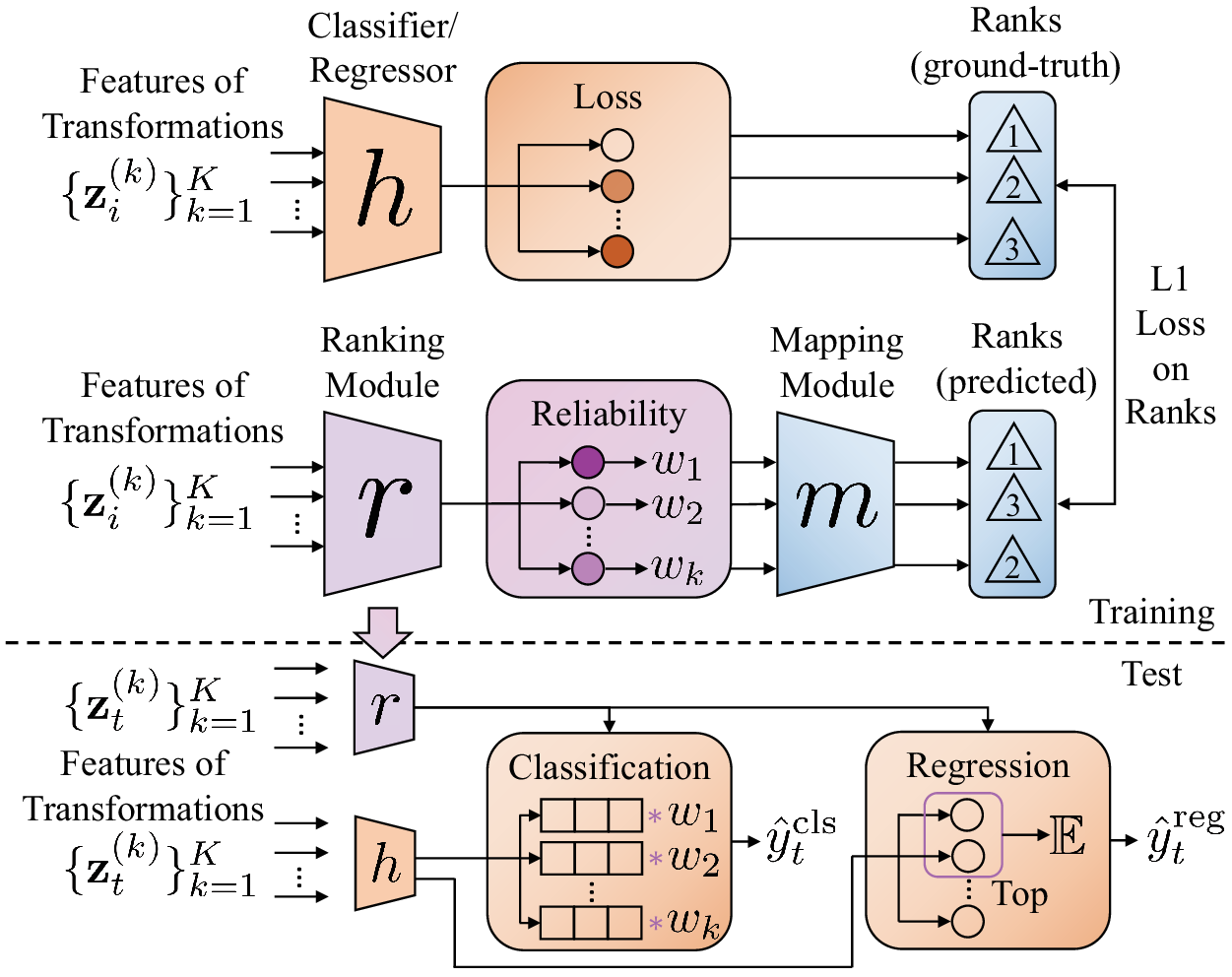}
\caption{Training and inference of the ranking module, and prediction aggregation strategy for classification and regression tasks, respectively.} \label{fig:learning2rank}
\end{figure}

To address these challenges, $r(\cdot)$ must output positively correlated reliability scores that ideally resemble discrete rankings. To this end, we adopt a learning-to-rank strategy~\cite{Engilberge2019} by introducing an auxiliary mapping module $m(\cdot)$, which transforms the task loss after Softmax normalization scores in $[0,1]$ space into a pseudo-discrete space $\{1, 2, \ldots, K\}$ representing rank-like values. Illustrations are shown in Fig.~\ref{fig:learning2rank}. Although $m(\cdot)$ produces continuous outputs, this transformation effectively amplifies the separation between similar reliability scores, thereby facilitating more accurate modeling of transformation quality.

Specifically, the mapping module $m(\cdot)$ is a light model that can be easily pre-trained on synthetic data. We followed~\cite{Engilberge2019} to generate synthetic samples $\mathcal{D}_{\text{synthetic}} = \{\tilde{\mathbf{x}}_{i}\}_{i=1}^{n_{\text{syn}}}$. Each synthetic sample is a vector $\tilde{\mathbf{x}}_i \in \mathbb{R}^K$ where each value of it is a randomly generated scalar $\tilde{x}_{i,k} \in [0,1]$. Its corresponding ground-truth rank vector is then $\tilde{\boldsymbol{\pi}}_i \in \mathbb{R}^K$ where each value is $\tilde{\pi}_{i,k} \in \{1, 2, \dots, K\}$. The optimization objective for $m(\cdot)$ is $L1$ loss, which is a standard metric for comparing two rankings:
\begin{align}
\mathcal{L}_{\text{mapping}}[m(\cdot)] = \mathbb{E}_{\tilde{\mathbf{x}}_i \sim \mathcal{D}_{\text{synthetic}}} \left\| m(\tilde{\mathbf{x}}_i) - \tilde{\boldsymbol{\pi}}_i \right\|_1. \label{eq:mappingloss}
\end{align}
Note that the input and output spaces of the mapping module is continuous, instead of discrete. This mapping module thus avoids non-differentiable projection into the ranking space.

After optimizing the mapping module $m(\cdot)$, the ranking module $r(\cdot)$ is trained using the training set $\{\mathbf{x}_i\}_{i=1}^{n_s}$. It takes features of transformations $\{\mathbf{z}_i^{(k)}\}_{k=1}^{K}$ from the pre-trained feature extractor $g(\cdot)$ as inputs, and outputs reliability scores. The mapping module then projects such scores into ranks. The ground-truth rank vector for the ranking of the transformations for a specific sample is determined by the task module. 

Specifically, the outputs of the ranking module $r(\cdot)$ first go through a Softmax function to convert into the weight vector $\mathbf{w}_i \in \mathbb{R}^{K}$ of values $w_{i,k}$ that sum up to one:
\begin{align}
w_{i,k} = \frac{\exp\left(r(\mathbf{z}_{i}^{(k)})\right)}{\sum_{j=1}^{K} \exp\left(r(\mathbf{z}_{i}^{(j)})\right)}. \label{eq:reliabilityweight}
\end{align}

The loss objective is therefore still a regression $L1$ loss in the integer-like space:
\begin{align}
\mathcal{L}_{\text{ranking}}[r(\cdot)] = \mathbb{E}_{\mathbf{x}_i \sim \mathcal{D}_{\text{train}}} \left\| m(\mathbf{w}_i) - \boldsymbol{\pi}_i \right\|_1. \label{eq:rankingloss}
\end{align}

To summarize, by decoupling transformation ranking from direct supervision via task loss, the learnable ranking module amplifies the distinction of reliability levels across transformations, handling the aforementioned limitations.

\subsection{Inference Aggregation}\label{sect:ensemble}

To aggregate the predictions to the multiple transformations, we define the following strategies:
\begin{itemize}
\item \textbf{Classification:} The core concept of ensemble for classification can be regarded as applying higher weights to more reliable predictions for a convex combination~\cite{Zhou2012}. The classifier's logit outputs $h(\mathbf{z}_{t}^{(k)})$ are first sharpened using temperature rescaling, a standard technique for adjusting the confidence of predictions prior to applying the Softmax function. The sharpened logits are then transformed into class probabilities via the Softmax function and aggregated using the reliability scores as weights from the ranking module $r(\cdot)$:
\begin{align}
\hat{y}_{t}^{\text{cls}}=\operatornamewithlimits{arg\,max}_{c \in \{1, \ldots, C\}} \bigg[ \sum_{k=1}^{K} w_{t,k} \frac{\exp\bigg( \left[ h(\mathbf{z}_{t}^{(k)}) \right]_c / \tau \bigg)}{\sum_{c'=1}^C \exp\bigg( \left[ h(\mathbf{z}_{t}^{(k)}) \right]_{c'} / \tau \bigg)} \bigg], \label{eq:clspred}
\end{align}
where $C$ denotes the number of classes, $w_{t,k}$ the reliability scores, and $\tau$ is the temperature hyperparameter. The value of $\tau$ is typically a power of two and less than one to ensure sharpening of the probability distribution.

\item \textbf{Regression:} For regression tasks, the weighted convex combination used for classification does not directly apply, since the outputs are continuous scalars rather than probability distributions over a simplex. We therefore aggregate by selecting the top-ranked half of the transformations, ordered by the reliability scores from $r(\cdot)$, and averaging their predictions. The objective in Eq.~\eqref{eq:rankingloss} constrains the relative order of the branch reliabilities rather than a calibrated inverse-error magnitude, so an order-based selection is the aggregation matched to the learned quantity. Given the order, the cutoff $\lceil K/2\rceil$ retains the more reliable majority of the branches. Let $k'_j$ denote the index of the transformation with the $j$-th highest value of $r(\mathbf{z}_t^{(k)})$. The aggregated prediction is:
\begin{align}
\hat{y}_{t}^{\text{reg}} = \frac{1}{\left\lceil \frac{K}{2} \right\rceil} \sum_{j=1}^{\left\lceil \frac{K}{2} \right\rceil} h\big(\mathbf{z}_{t}^{(k'_j)}\big), \label{eq:regpred}
\end{align}
where $\left\lceil \cdot \right\rceil$ denotes the ceiling operator.
\end{itemize}

\subsection{Summary of BFT}\label{sect:pseudocode}

The pseudo-code for BFT is presented in Algorithm~\ref{tab:alg}.

In summary, BFT reduces prediction uncertainty at the instance level, thereby implicitly addressing conditional distribution shifts, achieving gains similar to classic TL on uncertainty mitigation. Compared to classic approaches, BFT is backpropagation-free, privacy-preserving, noise-robust, and supports both classification and regression tasks. To address marginal distribution shifts, existing techniques such as EA~\cite{He2020EA, Li2024T-TIME} and BN-adapt~\cite{Schneider2020BNadapt} are effective and fully compatible with BFT, without conflicting with the core properties.

A variance-based theoretical analysis of this uncertainty reduction is provided in Section~S-I of the Supplementary Material. The theory identifies the conditions under which aggregating label-preserving transformations reduces prediction variance, and under which reliability weighting reduces it further when branch quality is heterogeneous.

\begin{algorithm}[htpb]
\caption{Backpropagation-Free Transformations (BFT)}
\label{tab:alg}
\begin{algorithmic}
\REQUIRE 
Streaming test data $\{\mathbf{x}_t\}_{t=1}^{n_t}$;\\
Labeled training data $\{(\mathbf{x}_i, y_i)\}_{i=1}^{n_s}$;\\
$g(\cdot)$, the trained feature extractor;\\
$h(\cdot)$, the trained classifier or regressor;\\
$r(\cdot)$, the ranking module;\\
$m(\cdot)$, the mapping module;\\
$K$, the number of transformations;\\
$\{\mathcal{T}_k(\cdot)\}_{k=1}^K$, the knowledge-guided transformation functions for BFT-A;\\
$\tau$, the temperature rescaling factor;\\
\ENSURE Prediction $\hat{y}_{t}^{\text{cls}}$ or $\hat{y}_{t}^{\text{reg}}$ for each $\mathbf{x}_t$;\\
\STATE \emph{// Mapping Module Training}
\STATE Generate $\mathcal{D}_{\text{synthetic}} = \{\tilde{\mathbf{x}}_i\}_{i=1}^{n_{\text{syn}}}$;\\
\STATE Compute ground-truth ranks $\{\tilde{\boldsymbol{\pi}}_i\}_{i=1}^{n_{\text{syn}}}$ for synthetic samples;\\
\STATE Train $m$ using $\mathcal{L}_{\text{mapping}}$ by Eq.~\eqref{eq:mappingloss};\\
\STATE \emph{// Ranking Module Training}
\STATE Calculate transformed features $\{\mathbf{z}_i^{(k)}\}_{i=1,k=1}^{n_s,K}$ using training data $\{\mathbf{x}_i\}_{i=1}^{n_s}$, by Eq.~\eqref{eq:BFT-A} for BFT-A, or by Eq.~\eqref{eq:dropoutmask} and Eq.~\eqref{eq:BFT-D} for BFT-D;\\
\STATE Compute ranking weights $\{w_{i,k}\}_{i=1}^{n_s}$ using Eq.~\eqref{eq:reliabilityweight};\\
\STATE Compute task-based rank labels $\{\boldsymbol{\pi}_i\}_{i=1}^{n_s}$;\\
\STATE Train $r$ using $\mathcal{L}_{\text{ranking}}$ by Eq.~\eqref{eq:rankingloss};\\
\STATE \emph{// Online Test Phase}
\FOR{$t = 1$ to $n_t$}
\STATE Calculate transformed features $\{\mathbf{z}_t^{(k)}\}_{k=1}^{K}$ using test input $\mathbf{x}_t$, by Eq.~\eqref{eq:BFT-A} for BFT-A, or by Eq.~\eqref{eq:dropoutmask} and Eq.~\eqref{eq:BFT-D} for BFT-D;\\
\IF{classification}
\STATE Compute classification prediction $\hat{y}_t^{\text{cls}}$ by Eq.~\eqref{eq:clspred};\\
\ELSIF{regression}
\STATE Compute regression prediction $\hat{y}_t^{\text{reg}}$ by Eq.~\eqref{eq:regpred}
\ENDIF
\ENDFOR
\end{algorithmic}
\end{algorithm}

\section{Experiments}\label{sect:experiments}

This section details the experiments that verified the effectiveness of BFT on EEG datasets. All algorithms were implemented in Python, and the code is available on GitHub\footnote{https://github.com/sylyoung/DeepTransferEEG}.

\subsection{Datasets}\label{sect:dataset}

A total of five EEG datasets were used in the experiments. Table~\ref{tab:datasets} summarizes the main characteristics of the datasets.

Three motor imagery (MI) EEG datasets were used under classification tasks. Subjects were asked to perform imagined body part movements for a few seconds. Different types of imaginary can be differentiated through the corresponding spatial sensorimotor rhythm modulations for BCI control. Left and right hand imagery tasks were considered.

Two driver-drowsiness estimation EEG datasets were used under regression tasks. EEG signals are used to estimate fatigue levels during driving (often simulated). Variations in neural patterns, such as increased theta or decreased alpha activity, reflect reduced vigilance~\cite{Fu2024, Du2023a}. For measuring fatigue levels of the subjects, reaction time was converted to drowsiness index~\cite{Cui2019} for the Driving dataset, while PERCLOS~\cite{Wierwille1994} was used for the SEED-VIG dataset. Both metrics range in $[0,1]$, with their calculation formulas available in the aforementioned publications. Thus, no further label normalization was applied.

\begin{table*}[htpb]  \centering
\caption{Summary of the five EEG datasets.}  \label{tab:datasets}
\begin{tabular}{c|c|c|c|c|c|c}
\toprule
\multirow{2}{*}{Dataset} & Number of & Number of & Sampling & Trial Length & Number of & Task \\
& Subjects & Channels & Rate (Hz) & (seconds) & Trials & Type \\
\midrule
Zhou2016~\cite{Zhou2016} & 4 & 14 & 250 & 5 & [90, 119] & left / right hand MI classification \\
BNCI2014001~\cite{Tangermann2012} & 9 & 22 & 250 & 4 & 144 & left / right hand MI classification \\
HighGamma~\cite{Schirrmeister2017} & 14 & 128 & 500 & 4 & [160, 448] & left / right hand MI classification \\
\midrule
Driving~\cite{Chuang2014} & 15 & 30 & 250 & 8 & [1015, 1197] & reaction time in drowsiness index [0, 1] regression \\
SEED-VIG~\cite{Zheng2017} & 23 & 17 & 200 & 8 & 885 & PERCLOS [0, 1] regression \\
\bottomrule
\end{tabular}
\end{table*}

\subsection{Experiment Settings} \label{sect:settings}  

We considered a plug-and-play evaluation setting under leave-one-subject-out cross-validation. For each experiment, one subject's data was held out as the test set, while data from the remaining subjects were combined as the training set. No information from the test set was accessible during the training phase, and the test phase was conducted using ordered trial-wise online data streams. Only the first session data were used to focus the study on inter-subject discrepancies.

All experiments were repeated three times with different random seeds. Since the used datasets contained many subjects, we report dataset-wise averaged performance scores (except for Zhou2016, which reported subject-wise scores), with standard deviations of variations across repeated experiments.

Classification performance was evaluated using accuracy, while regression performance was evaluated using the Pearson correlation coefficient (CC) and root mean squared error (RMSE) metrics.

To mitigate marginal distribution shift, we employed EA~\cite{He2020EA, Li2024T-TIME} and BN-adapt~\cite{Schneider2020BNadapt}, which are effective, backpropagation-free, and computationally efficient. These methods were integrated into all TTA approaches.

The backbone architecture used was EEGNet~\cite{Lawhern2018EEGNet}, a lightweight convolutional neural network architecture for EEG decoding. $g$ denotes the convolutional layers of EEGNet, $h$ a fully-connected layer. $\tau$ was set to $0.5$.

The ranking module $r(\cdot)$ is a fully-connected network. The bi-directional long short-term memory (Bi-LSTM)~\cite{Schuster1997} is used only within the auxiliary mapping module $m(\cdot)$, which processes a length-$K$ vector of branch reliability scores rather than the temporal samples of an EEG trial. The branch positions are fixed only for tensor alignment, and the synthetic training values are assigned randomly across all positions, so the module encodes no temporal order and no preferred transformation. Following~\cite{Engilberge2019}, the synthetic score vectors for training $m(\cdot)$ were drawn from:
\begin{enumerate}
\item A uniform distribution over the interval $[-1, 1]$;
\item A normal distribution with mean $\mu = 0$ and standard deviation $\sigma = 1$;
\item A sequence of evenly spaced numbers within a uniformly drawn random sub-range of $[-1, 1]$;
\item Random mixtures of the above distributions.
\end{enumerate}

\subsection{EEG Transformations} \label{sect:transformations}

The following transformations were applied to EEG trials during the experiments, most of which were introduced in Section~\ref{sect:TTT}. Every branch presented exactly $t-1$ seconds to the model, where $t$ is the recorded trial duration. The identity, scaling, noise, and frequency-shift branches used the common reference crop $[0,t-1]$, while the sliding branches used equal-length crops at shifted onsets, so no branch used more temporal samples than another. The shifted onsets act as deliberate onset perturbations that probe robustness to trial-timing variation, consistent with the uncertainty-suppression principle of BFT.
\begin{enumerate}
\item Identity: The common reference crop $[0,t-1]$ is used without signal modification.
\item Amplitude Scaling (Scale): Each trial is scaled by one of the following factors: $\{0.9, 1.1, 1.2|$.
\item Noise Addition (Noise): Gaussian noise proportional to the signal magnitude of each channel is added.
\item Frequency Shift (Freq): Low- and high-frequency components are selectively shifted.
\item Sliding Window (Slide): Five temporal segments of duration $t-1$ are cropped from the recorded trial: $[0.2, t-0.8]$, $[0.4, t-0.6]$, $[0.6, t-0.4]$, $[0.8, t-0.2]$, and $[1, t]$. These equal-length views simulate variations in signal onset relative to the reference crop.
\item Channel Reflection~\cite{Wang2024} and Discrete Wavelet Transform Augmentation~\cite{Wang2025}: These enhancements are label-aware transformations, and thus are applied only during training to improve the performance of the task module $h(g(\cdot))$ in classification tasks.
\end{enumerate}

In total, fourteen transformations were used during training for classification tasks and twelve for regression tasks. On-the-fly augmentation was adopted, where each training sample was randomly transformed using one of the augmentation techniques with equal probability in each epoch. During test-time transformations in BFT-A, $K=12$ types of transformations were applied to each test trial. $K=10$ was used for BFT-D.

\subsection{Results for Classification Task} \label{sect:classification}

The following approaches were evaluated:
\begin{enumerate}
\item CSP-LDA: Common Spatial Pattern filters were used for feature extraction, followed by Linear Discriminant Analysis. Repeated experiments use 5, 6, and 7 CSP filters.
\item EEGNet: The backbone is trained with cross-entropy loss, with or without data augmentation. The augmented EEGNet checkpoint is used as the common source model for all TTA methods to ensure a fair comparison.
\item UDA: DAN~\cite{Long2019}, JAN~\cite{Long2017}, DANN~\cite{Ganin2016}, CDAN-E~\cite{Long2018}, MDD~\cite{Zhang2019}, MCC~\cite{Jin2024}, and SHOT-IM~\cite{Liang2022} adapt the source representation or decision boundary using unlabeled target data.
\item TTA with backpropagation: MEMO, Tent, PL, SAR, and T-TIME update model parameters from unlabeled test samples.
\item TTA without backpropagation: BN-adapt, T3A~\cite{Iwasawa2021}, and LAME~\cite{Boudiaf2022LAME} adapt predictions or feature statistics without gradient-based parameter updates.
\item Transformation-based TTA: Aug-Scale, Aug-Noise, Aug-Freq, and Aug-Slide apply one transformation family at inference. Aug-Mean uniformly averages the augmented views, whereas Mask-Mean uniformly averages the masked branches. They are the unweighted counterparts of BFT-A and BFT-D, respectively.
\end{enumerate}

The classification accuracies on the three MI datasets are shown in Tables~\ref{tab:Zhou2016} and~\ref{tab:BNCIandHG}. Observe that:

\begin{table}[htpb] \centering
\caption{Subject-wise cross-subject classification accuracy (\%) on Zhou2016. Bold denotes the category best. $^{***}$ marks a BFT variant significantly above its unweighted counterpart under a two-sided paired Wilcoxon test: $p=9\times10^{-4}$ for both BFT-A versus Aug-Mean and BFT-D versus Mask-Mean (12 seed-subject pairs).}  \label{tab:Zhou2016}
\scalebox{0.87}{\begin{tabular}{c|c|c|c|c|c|c}
\toprule
Category & Approach & S1 & S2 & S3 & S4 &  Avg. \\
\midrule
\multirow{3}{*}{\shortstack{w/o \\ TL}} & CSP-LDA & 72.55 & 77.33 & 88.00 & 82.22 & 80.03$_{\pm0.56}$ \\
~ & EEGNet (w/o Aug.) & 82.35 & 75.33 & 89.00 & 80.74 & 81.86$_{\pm0.98}$ \\
~ & EEGNet & 80.95  & 81.00  & 93.67  & 79.63  & \textbf{83.81}$_{\pm2.06}$ \\
\midrule
\multirow{7}{*}{UDA} & DAN & 78.43 & 78.67 & 89.33 & 75.56 & 80.50$_{\pm2.02}$ \\
~ & JAN & 78.43 & 78.67 & 87.67 & 80.37 & 81.29$_{\pm2.50}$ \\
~ & DANN & 78.15 & 78.00 & 89.33 & 77.41 & 80.72$_{\pm0.83}$ \\
~ & CDAN-E & 78.43 & 78.00 & 89.67 & 82.96 & 82.27$_{\pm2.03}$ \\
~ & MDD & 78.71 & 77.67 & 90.67 & 74.81 & 80.47$_{\pm0.90}$ \\
~ & MCC & 82.91 & 81.67 & 93.00 & 90.00 & \textbf{86.90}$_{\pm0.23}$ \\
~ & SHOT-IM & 82.91  & 80.00  & 94.00  & 84.07  & 85.25$_{\pm1.47}$ \\
\midrule
\multirow{5}{*}{\shortstack{TTA \\ w/ \\ BP}} & MEMO & 81.79  & 82.33  & 94.00  & 81.11  & 84.81$_{\pm2.26}$ \\
~ & Tent & 80.39  & 76.67  & 93.00  & 81.11  & 82.79$_{\pm2.11}$ \\
~ & PL & 83.19  & 77.00  & 93.67  & 85.18  & 84.76$_{\pm2.32}$ \\
~ & SAR & 80.67  & 73.00  & 92.33  & 85.18  & 82.80$_{\pm0.96}$ \\
~ & T-TIME & 83.75  & 78.00  & 93.33  & 86.30  & \textbf{85.35}$_{\pm0.82}$ \\
\midrule
\multirow{11}{*}{\shortstack{TTA \\ w/o \\ BP}} & BN-adapt & 82.35  & 79.00  & 94.00  & 80.37  & 83.93$_{\pm1.34}$ \\
~ & T3A & 73.95  & 74.67  & 91.00  & 56.30  & 73.98$_{\pm1.71}$ \\
~ & LAME & 84.03 & 77.33 & 93.33 & 79.26 & 83.49$_{\pm1.31}$ \\ \cmidrule{2-7}
~ & Aug-Scale & 80.95  & 80.00  & 93.67  & 79.26  & 83.47$_{\pm1.44}$ \\
~ & Aug-Noise & 80.95  & 80.67  & 92.67  & 80.00  & 83.57$_{\pm1.59}$ \\
~ & Aug-Freq & 80.95  & 80.33  & 93.67  & 80.37  & 83.83$_{\pm2.62}$ \\
~ & Aug-Slide & 82.35  & 77.33  & 93.00  & 80.37  & 83.26$_{\pm0.88}$ \\
~ & Mask-Mean & 80.95  & 81.00  & 93.67  & 79.63  & 83.81$_{\pm2.06}$ \\
~ & Aug-Mean & 82.91  & 78.00  & 93.67  & 80.37  & 83.74$_{\pm2.67}$ \\ \cmidrule{2-7}
~ & BFT-D (ours) & 82.63  & 79.33  & 93.33  & 82.22  & 84.38$_{\pm1.22}^{***}$ \\
~ & BFT-A (ours) & 84.03  & 78.00  & 94.33  & 84.08  & \textbf{85.11}$_{\pm1.27}^{***}$ \\
\bottomrule
\end{tabular}}
\end{table}

\begin{table}[htpb] \centering
\caption{Dataset-wise cross-subject classification accuracy (\%) on BNCI2014001 and HighGamma. Bold denotes the category best. $^{***}$ marks a BFT variant significantly above its unweighted counterpart under a two-sided paired Wilcoxon test: on BNCI2014001, $p=6\times10^{-6}$ for BFT-A versus Aug-Mean and $p=1\times10^{-5}$ for BFT-D versus Mask-Mean (27 seed-subject pairs); on HighGamma, $p=2\times10^{-7}$ and $p=4\times10^{-7}$ for the same two comparisons (42 seed-subject pairs).}  \label{tab:BNCIandHG}
\scalebox{1.0}{\begin{tabular}{c|c|c|c}
\toprule
Category & Approach & BNCI2014001 & HighGamma \\
\midrule
\multirow{3}{*}{\shortstack{w/o \\ TL}} & CSP-LDA & 72.76$_{\pm0.31}$ & 67.46$_{\pm1.02}$\\
~ & EEGNet (w/o Aug.) & 75.39$_{\pm1.22}$ & 74.03$_{\pm0.61}$\\
~ & EEGNet & \textbf{76.49}$_{\pm0.45}$ & \textbf{77.55}$_{\pm0.26}$\\
\midrule
\multirow{7}{*}{UDA} & DAN & 77.24$_{\pm0.98}$ & 75.42$_{\pm0.88}$\\
~ & JAN & 74.90$_{\pm1.11}$ & 74.04$_{\pm0.10}$ \\
~ & DANN & 75.59$_{\pm1.73}$ & 75.41$_{\pm1.05}$ \\
~ & CDAN-E & 78.76$_{\pm1.66}$ & 73.94$_{\pm0.46}$ \\
~ & MDD & 76.44$_{\pm1.10}$ & 75.43$_{\pm0.16}$ \\
~ & MCC & \textbf{79.91}$_{\pm1.12}$ & 66.25$_{\pm0.97}$ \\
~ & SHOT-IM & 79.22$_{\pm0.27}$ & \textbf{77.72}$_{\pm0.47}$\\
\midrule
\multirow{5}{*}{\shortstack{TTA \\ w/ \\ BP}} & MEMO & 76.80$_{\pm0.37}$ & \textbf{78.19}$_{\pm0.34}$\\
~ & Tent & 74.56$_{\pm1.29}$ & 71.61$_{\pm1.73}$ \\
~ & PL & 77.13$_{\pm1.55}$ & 76.00$_{\pm1.84}$ \\
~ & SAR & 77.37$_{\pm0.48}$ & 71.64$_{\pm2.00}$ \\
~ & T-TIME & \textbf{79.22}$_{\pm0.80}$ & 77.42$_{\pm0.76}$ \\
\midrule
\multirow{11}{*}{\shortstack{TTA \\ w/o \\ BP}} & BN-adapt & 76.94$_{\pm0.43}$ & 78.23$_{\pm0.45}$ \\
~ & T3A & 69.75$_{\pm3.45}$ & 61.10$_{\pm1.40}$ \\
~ & LAME & 75.41$_{\pm1.09}$ & 77.74$_{\pm0.35}$ \\ \cmidrule{2-4}
~ & Aug-Scale & 76.34$_{\pm0.38}$ & 77.56$_{\pm0.40}$ \\
~ & Aug-Noise & 76.21$_{\pm0.51}$ & 77.72$_{\pm0.23}$ \\
~ & Aug-Freq & 76.13$_{\pm1.13}$ & 77.21$_{\pm0.70}$ \\
~ & Aug-Slide & 69.81$_{\pm1.62}$ & 75.65$_{\pm0.68}$ \\
~ & Mask-Mean & 76.52$_{\pm0.48}$ & 77.55$_{\pm0.26}$ \\
~ & Aug-Mean & 76.31$_{\pm0.60}$ & 78.09$_{\pm0.68}$ \\ \cmidrule{2-4}
~ & BFT-D (ours) & 77.47$_{\pm0.54}^{***}$ & 78.54$_{\pm0.40}^{***}$ \\
~ & BFT-A (ours) & \textbf{77.80}$_{\pm0.96}^{***}$ & \textbf{79.03}$_{\pm0.43}^{***}$ \\
\bottomrule
\end{tabular}}
\end{table}

\begin{enumerate}
\item Performance of individual transformations were inconsistent across datasets: a transformation that helped on one dataset could fall well below the source model on another, whereas the uniform aggregations (Aug-Mean and Mask-Mean) were more stable. This confirms that no single transformation is reliably beneficial and motivates reliability-aware aggregation of BFT.
\item Both BFT variants outperformed their unweighted counterparts on all three datasets, i.e., BFT-A over Aug-Mean and BFT-D over Mask-Mean, and every improvement was statistically significant under two-sided paired Wilcoxon tests. This isolates the contribution of the learned ranking beyond simple averaging.
\item BFT-A was the strongest backpropagation-free approach on all three datasets, and its accuracy was comparable to the best approach under TTA w/ BP, despite using only forward passes. It surpassed the state-of-the-art T-TIME on HighGamma, while remaining slightly below it on BNCI2014001.
\item This dataset-dependent contrast is consistent with the spatial sampling density. HighGamma records many more channels than BNCI2014001, so its denser montage carries more redundant spatial information, from which the knowledge-guided views and structured masks generate diverse yet label-preserving branches that aggregate more effectively. Low-channel data offer less spatial redundancy for BFT to exploit, and thus favor explicit parameter adaptation.
\end{enumerate}

To test whether the gains depend on EEGNet, we additionally replaced the backbone with EEG Conformer~\cite{Song2023Conformer}, a convolutional-transformer hybrid that couples a convolutional front-end with a multi-head self-attention encoder and is substantially larger and deeper than the lightweight EEGNet. Table~\ref{tab:conformer} shows that BFT-A and BFT-D improved over Conformer (with data augmentation) on all three MI datasets. This indicates that the reliability-aware aggregation of BFT transfers across backbones, including a self-attention architecture, rather than exploiting a property specific to EEGNet.

\begin{table}[htpb]
\centering
\caption{Dataset-wise cross-subject classification accuracy (\%) with the EEG Conformer backbone. Bold denotes the row best.}
\label{tab:conformer}
\begin{tabular}{l|c|cc}
\toprule
Dataset & Conformer & +BFT-D & +BFT-A \\
\midrule
Zhou2016 & 81.41$_{\pm0.57}$ & 82.79$_{\pm0.84}$ & \textbf{83.26}$_{\pm1.00}$ \\
BNCI2014001 & 69.65$_{\pm0.16}$ & 70.68$_{\pm0.45}$ & \textbf{71.09}$_{\pm0.63}$ \\
HighGamma & 76.24$_{\pm0.39}$ & 77.52$_{\pm0.22}$ & \textbf{77.82}$_{\pm0.24}$ \\
\bottomrule
\end{tabular}
\end{table}

\subsection{Results for Regression Task} \label{sect:regression}

Most existing TTA methods are designed for classification and rely on entropy, class probabilities, class-balanced priors, or pseudo-class labels that do not directly extend to continuous targets. We therefore evaluated the following regression-compatible approaches:
\begin{enumerate}
\item PSD-MLP~\cite{Cui2019}: Power Spectral Density features are extracted and passed to a multilayer perceptron regressor.
\item EEGNet: The same compact backbone is trained with mean squared error, with or without data augmentation. The augmented model is the common source checkpoint for test-time methods.
\item UDA: DAN, DANN, CORAL~\cite{Sun2016}, and DARE-GRAM~\cite{Nejjar2023} provide regression-compatible source-target adaptation baselines.
\item TTA without backpropagation: BN-adapt updates batch-normalization statistics; Aug-Scale, Aug-Noise, Aug-Freq, and Aug-Slide apply individual test-time transformations; Aug-Mean and Mask-Mean are the corresponding unweighted aggregation baselines for BFT-A and BFT-D.
\end{enumerate}

Table~\ref{tab:regression} shows that both BFT variants improved the correlation coefficient and reduced the RMSE over their unweighted aggregation baselines, Aug-Mean and Mask-Mean, on both driver-drowsiness datasets, while remaining competitive with the UDA baselines that additionally require source data.

The absolute gains are smaller than in classification because of the different output geometry. In classification, the sharpened probability vectors of different branches can move the weighted aggregate across a decision boundary, whereas in regression each branch produces a smoothly varying scalar, so a convex aggregation stays within a narrow range and there is less branch disagreement to exploit. As formalized in Section~\ref{sect:ensemble}, the top-half selection is the aggregation matched to the ordinal reliability that $r(\cdot)$ supervises for continuous outputs. Even so, the simultaneous correlation increases and RMSE reductions confirm that BFT consistently suppresses less reliable transformed predictions in continuous-output tasks.

\begin{table}[htpb] \centering
\caption{Dataset-wise cross-subject regression CCs and RMSEs on the two driver-drowsiness estimation EEG datasets. The best score for each category is marked in bold.}  \label{tab:regression}
\setlength{\tabcolsep}{3pt}
\resizebox{\linewidth}{!}{\begin{tabular}{c|c|cc|cc}
\toprule
\multirow{2.5}{*}{Category} & \multirow{2.5}{*}{Approach} & \multicolumn{2}{c|}{Driving} & \multicolumn{2}{c}{SEED-VIG} \\ \cmidrule{3-4} \cmidrule{5-6}
& & CC $\uparrow$ & RMSE $\downarrow$ & CC $\uparrow$ & RMSE $\downarrow$ \\ 
\midrule
\multirow{3}{*}{\shortstack{w/o \\ TL}} & PSD-MLP & 0.345$_{\pm0.033}$ & 0.546$_{\pm0.083}$ & 0.373$_{\pm0.007}$ & 0.331$_{\pm0.049}$ \\
~ & EEGNet (w/o Aug.) & \textbf{0.516}$_{\pm0.011}$ & \textbf{0.275}$_{\pm0.001}$ & \textbf{0.618}$_{\pm0.002}$ & 0.225$_{\pm0.004}$ \\
~ & EEGNet & 0.504$_{\pm0.017}$ & 0.276$_{\pm0.004}$ & 0.618$_{\pm0.006}$ & \textbf{0.223}$_{\pm0.003}$ \\
\midrule
\multirow{4}{*}{UDA} & DAN & 0.522$_{\pm0.018}$ & 0.272$_{\pm0.008}$ & 0.609$_{\pm0.011}$ & 0.216$_{\pm0.003}$ \\
~ & DANN & 0.530$_{\pm0.008}$ & 0.269$_{\pm0.006}$ & \textbf{0.612}$_{\pm0.008}$ & 0.213$_{\pm0.003}$ \\
~ & CORAL & \textbf{0.531}$_{\pm0.005}$ & \textbf{0.264}$_{\pm0.003}$ & 0.611$_{\pm0.006}$ & \textbf{0.209}$_{\pm0.003}$ \\
~ & DARE-GRAM & 0.511$_{\pm0.008}$ & 0.275$_{\pm0.008}$ & 0.609$_{\pm0.009}$ & 0.215$_{\pm0.003}$ \\
\midrule
\multirow{9}{*}{\shortstack{TTA \\ w/o \\ BP}} 
& BN-adapt & 0.526$_{\pm0.010}$ & 0.278$_{\pm0.008}$ & 0.618$_{\pm0.010}$ & 0.216$_{\pm0.004}$ \\ \cmidrule{2-6}
~ & Aug-Scale & 0.506$_{\pm0.018}$ & 0.275$_{\pm0.004}$ & 0.619$_{\pm0.005}$ & 0.223$_{\pm0.003}$ \\
~ & Aug-Noise & 0.502$_{\pm0.016}$ & 0.275$_{\pm0.003}$ & 0.618$_{\pm0.006}$ & 0.222$_{\pm0.002}$ \\
~ & Aug-Freq & 0.504$_{\pm0.017}$ & 0.276$_{\pm0.005}$ & 0.617$_{\pm0.006}$ & 0.223$_{\pm0.002}$ \\
~ & Aug-Slide & 0.504$_{\pm0.018}$ & 0.276$_{\pm0.004}$ & 0.618$_{\pm0.004}$ & 0.223$_{\pm0.003}$ \\
~ & Mask-Mean & 0.504$_{\pm0.017}$ & 0.278$_{\pm0.004}$ & 0.618$_{\pm0.006}$ & 0.218$_{\pm0.001}$ \\
~ & Aug-Mean & 0.510$_{\pm0.017}$ & 0.277$_{\pm0.001}$ & 0.625$_{\pm0.005}$ & 0.222$_{\pm0.003}$ \\
\cmidrule{2-6}
~ & BFT-D (ours) & 0.534$_{\pm0.009}$ & 0.272$_{\pm0.005}$ & 0.623$_{\pm0.007}$ & \textbf{0.207}$_{\pm0.002}$ \\
~ & BFT-A (ours) & \textbf{0.535}$_{\pm0.008}$ & \textbf{0.271}$_{\pm0.006}$ & \textbf{0.629}$_{\pm0.005}$ & 0.208$_{\pm0.002}$ \\
\bottomrule
\end{tabular}}
\end{table}

\subsection{Test-Time Robustness} \label{sect:robust}

\begin{figure*}[htpb]\centering
\includegraphics[width=\linewidth,clip]{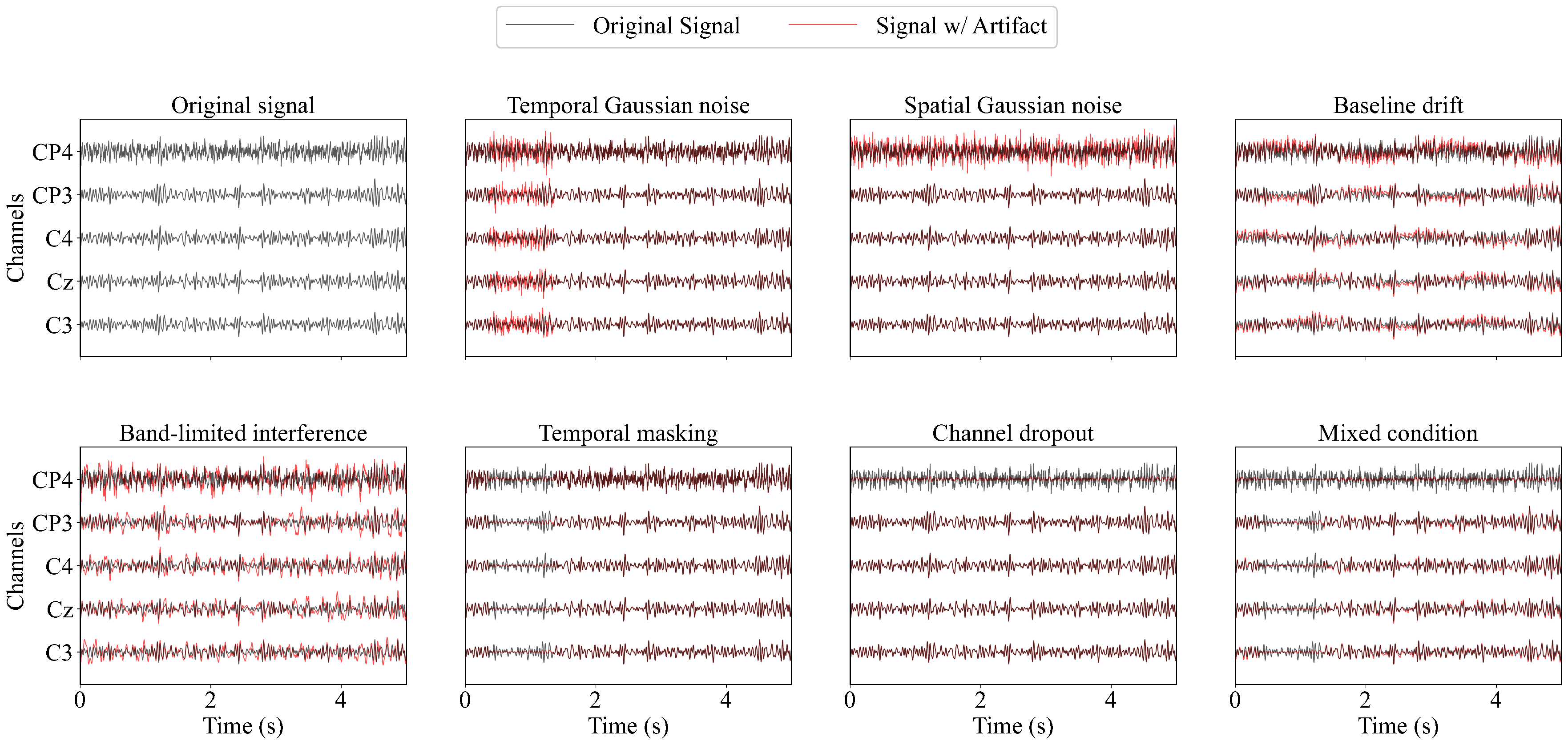}
\caption{Test-time corruptions applied to one EEG trial from Zhou2016 (five central channels C3, Cz, C4, CP3, CP4). In each panel the original signal is drawn in black and the corrupted signal in red. The panels show the original signal, the two broad Gaussian noise types (temporal and spatial), and five structured artifacts beyond Gaussian noise: baseline drift, band-limited interference, temporal masking, channel dropout, and a mixed condition.} \label{fig:noise}
\end{figure*}

This subsection investigates the robustness of TTA approaches to unexpected test-time corruption. As discussed in Section~\ref{sect:ttadeficiency}, practical EEG-based BCIs inevitably encounter signal contamination that degrades the quality of test samples. We simulate seven such corruptions and inject them into test trials, all illustrated in Fig.~\ref{fig:noise}. Two are Gaussian noise types:
\begin{enumerate}
\item Temporal noise, resulting from factors such as body movements. To simulate this, Gaussian noise was added to the temporal segment between $[1.5, 2.0]$ seconds of each test trial, with variance proportional to the signal magnitude for each channel.
\item Spatial noise, resulting from poor electrode-skin contact, etc. This is simulated by injecting Gaussian noise again into a single random channel over the entire trial duration, with variance proportional to the signal magnitude of that specific channel.
\end{enumerate}
The remaining five are structured artifacts that reproduce specific acquisition failures rather than diffuse noise: baseline drift, band-limited interference, temporal masking, channel dropout, and a mixed condition that combines several at once. All seven can be regarded as transformation functions.

\begin{figure*}[htpb]
\centering
\subfigure[]{\includegraphics[width=0.49\linewidth,clip]{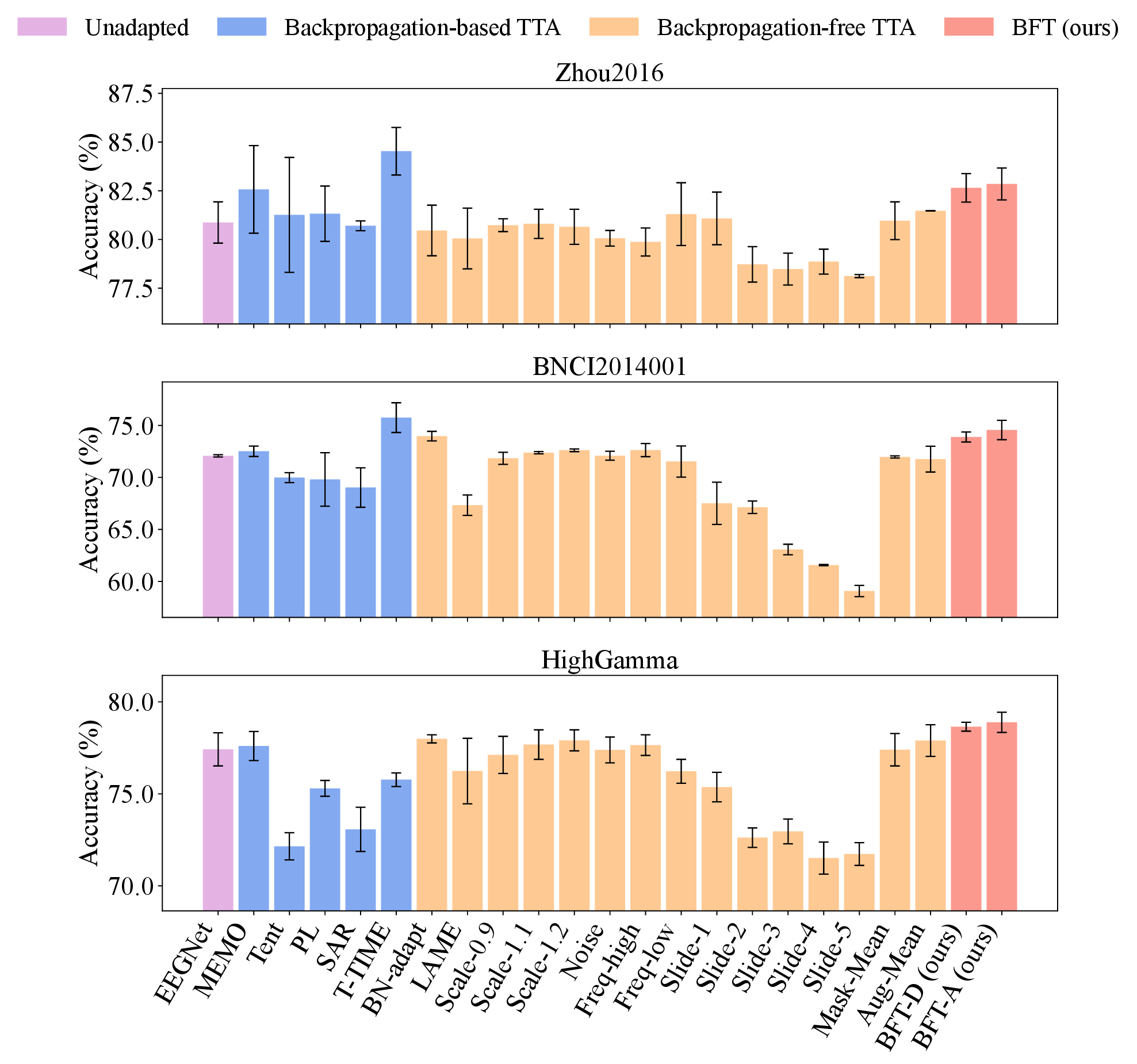}\label{fig:noise_classification_temporal}}
\hfill
\subfigure[]{\includegraphics[width=0.49\linewidth,clip]{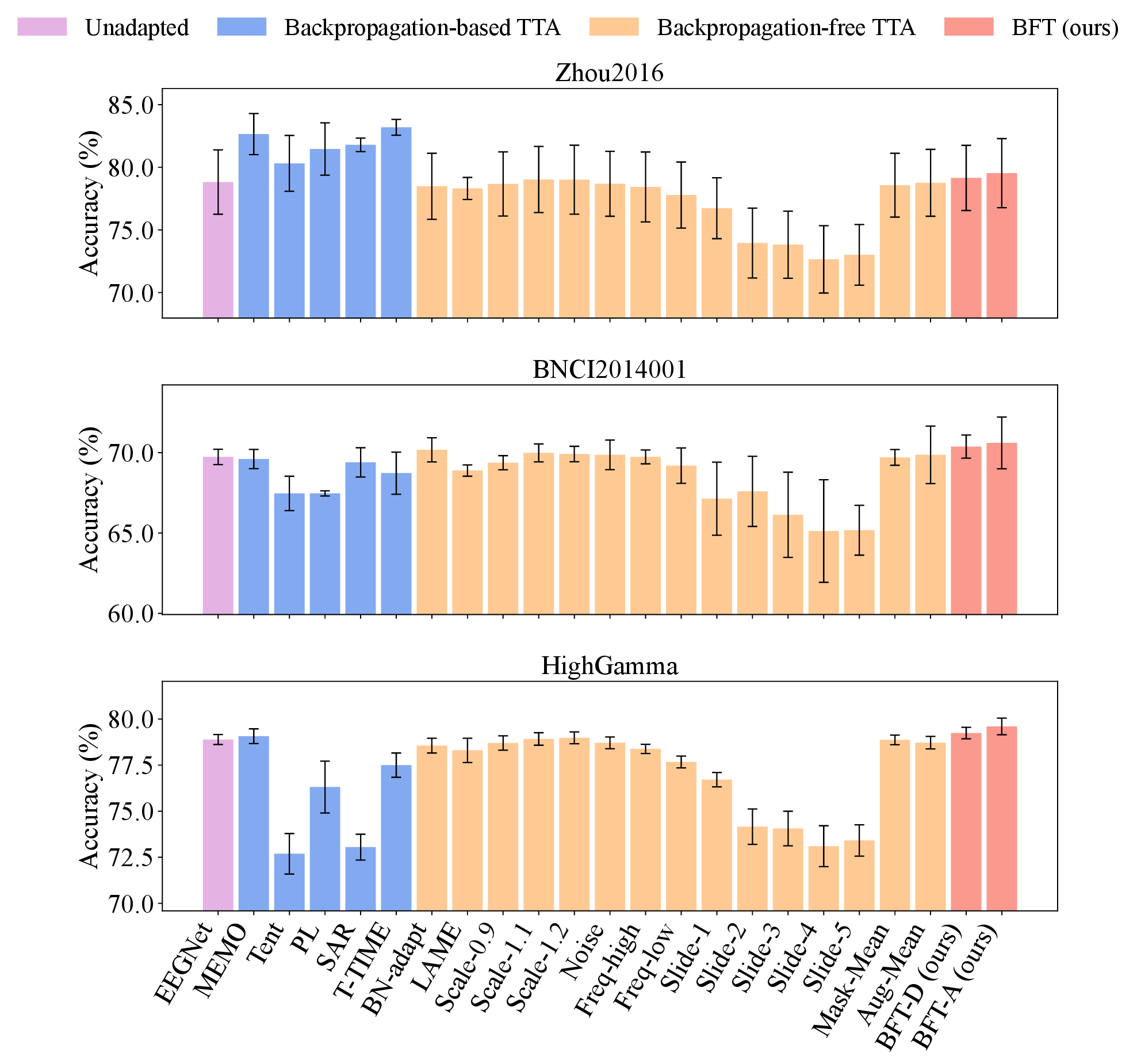}\label{fig:noise_classification_spatial}}
\caption{Accuracy (\%) under temporal and spatial noise during test phase for the three MI classification datasets. (a) temporal noise; and (b) spatial noise.}
\label{fig:noise_classification}
\end{figure*}

\begin{figure*}[htpb]
\centering
\subfigure[]{\includegraphics[width=0.49\linewidth,clip]{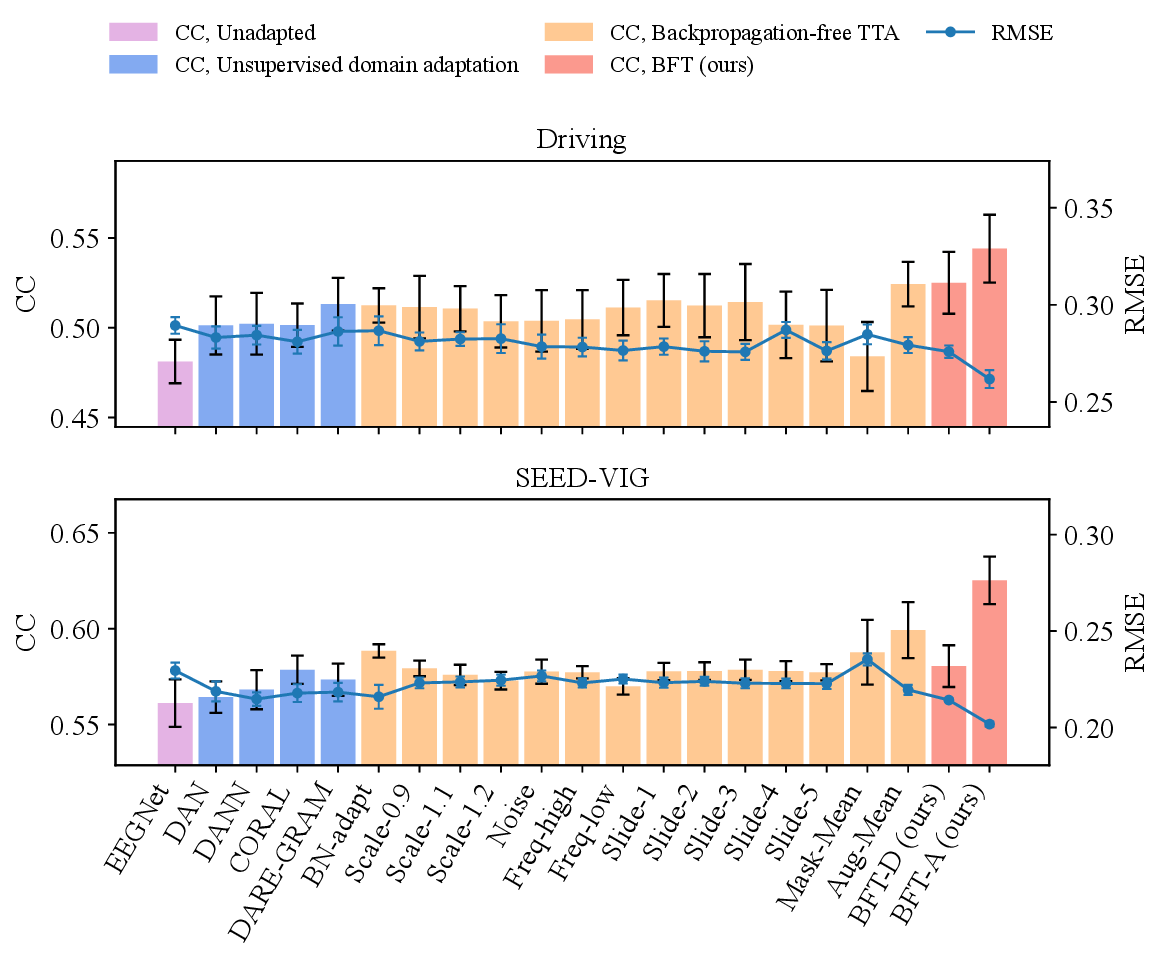}\label{fig:noise_regression_temporal}}
\hfill
\subfigure[]{\includegraphics[width=0.49\linewidth,clip]{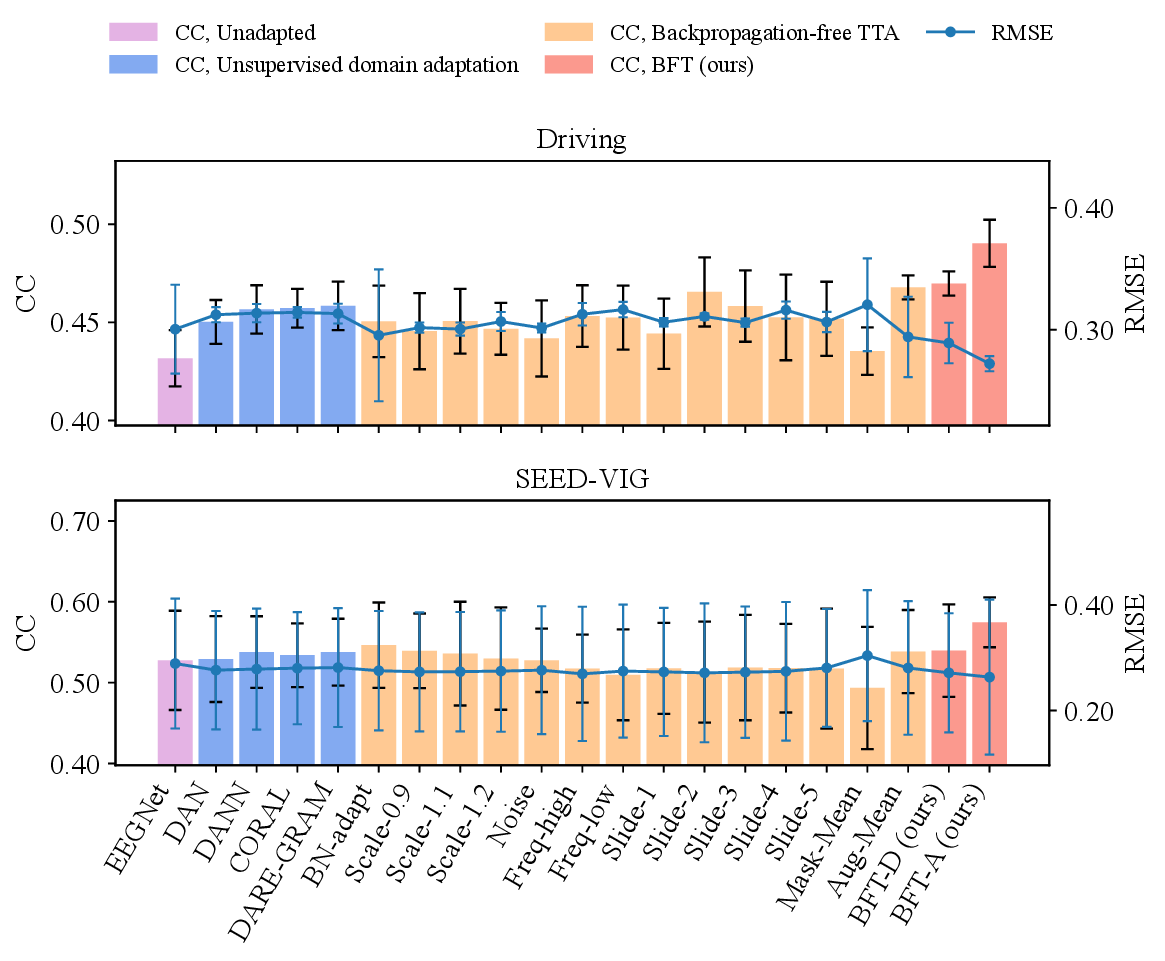}\label{fig:noise_regression_spatial}}
\caption{CCs and RMSEs under temporal and spatial noise during test phase for the two driver-drowsiness regression datasets. (a) temporal noise; and (b) spatial noise.}
\label{fig:noise_regression}
\end{figure*}

The results are presented in Fig.~\ref{fig:noise_classification} and Fig.~\ref{fig:noise_regression}. Observe that:
\begin{enumerate}
\item Under temporal noise, both BFT-A and BFT-D generally maintained their original performance across all five datasets, whereas the baseline and other TL approaches suffered different extents of performance drop.
\item Under spatial noise, all approaches suffered a drop in the absolute metric values, along with markedly higher instability. Nevertheless, BFT-A and BFT-D still achieved the best performance in all cases for the forward-only TTA family. Spatial noise is thus more challenging to address, likely because both paradigms depend heavily on spatial information, which the EEGNet architecture also emphasizes.
\end{enumerate}

\subsection{Ablation Studies}\label{sect:ablation}

We first compared the full model with two variants that train $r(\cdot)$ directly from real task losses without $m(\cdot)$: Variant~1 uses inverse predicted losses as weights, whereas Variant~2 converts them to integer ranks. As Table~\ref{tab:mappingablation} shows, the full BFT-A obtained the highest mean accuracy, and both full variants had markedly lower variability across subjects. For BFT-D, the full model was within noise of the best variant in mean accuracy but far more stable. The mapping module thus improves and stabilizes the reliability ranking, rather than uniformly maximizing every mean.

\begin{table}[htpb] \centering
\caption{Subject-wise cross-subject classification accuracy (\%) on Zhou2016. Bold denotes the highest mean accuracy of each of BFT-D and BFT-A across the three ranking schemes (Variant~1, Variant~2, and the full BFT).} \label{tab:mappingablation}
\begin{tabular}{c|c|c|c|c|c|c}
\toprule
Category & Approach & S1 & S2 & S3 & S4 & Avg. \\
\midrule
\multirow{2}{*}{\shortstack{Variant 1}} & BFT-D & 82.63 & 79.00 & 93.67 & 82.59 & \textbf{84.47}$_{\pm2.28}$ \\
~ & BFT-A & 83.75 & 77.67 & 93.67 & 83.33 & 84.60$_{\pm2.56}$ \\
\midrule
\multirow{2}{*}{\shortstack{Variant 2}} & BFT-D & 82.35 & 79.33 & 93.67 & 81.85 & 84.30$_{\pm2.33}$ \\
~ & BFT-A & 83.75 & 77.33 & 93.00 & 82.59 & 84.17$_{\pm2.33}$ \\
\midrule
\multirow{2}{*}{\shortstack{BFT}} & BFT-D & 82.62 & 79.33 & 93.33 & 82.22 & 84.38$_{\pm1.22}$ \\
~ & BFT-A & 84.03 & 78.00 & 94.33 & 84.08 & \textbf{85.11}$_{\pm1.27}$ \\
\bottomrule
\end{tabular}
\end{table}

\begin{figure*}[htpb] \centering
\subfigure[]{\includegraphics[width=.47\linewidth,clip]{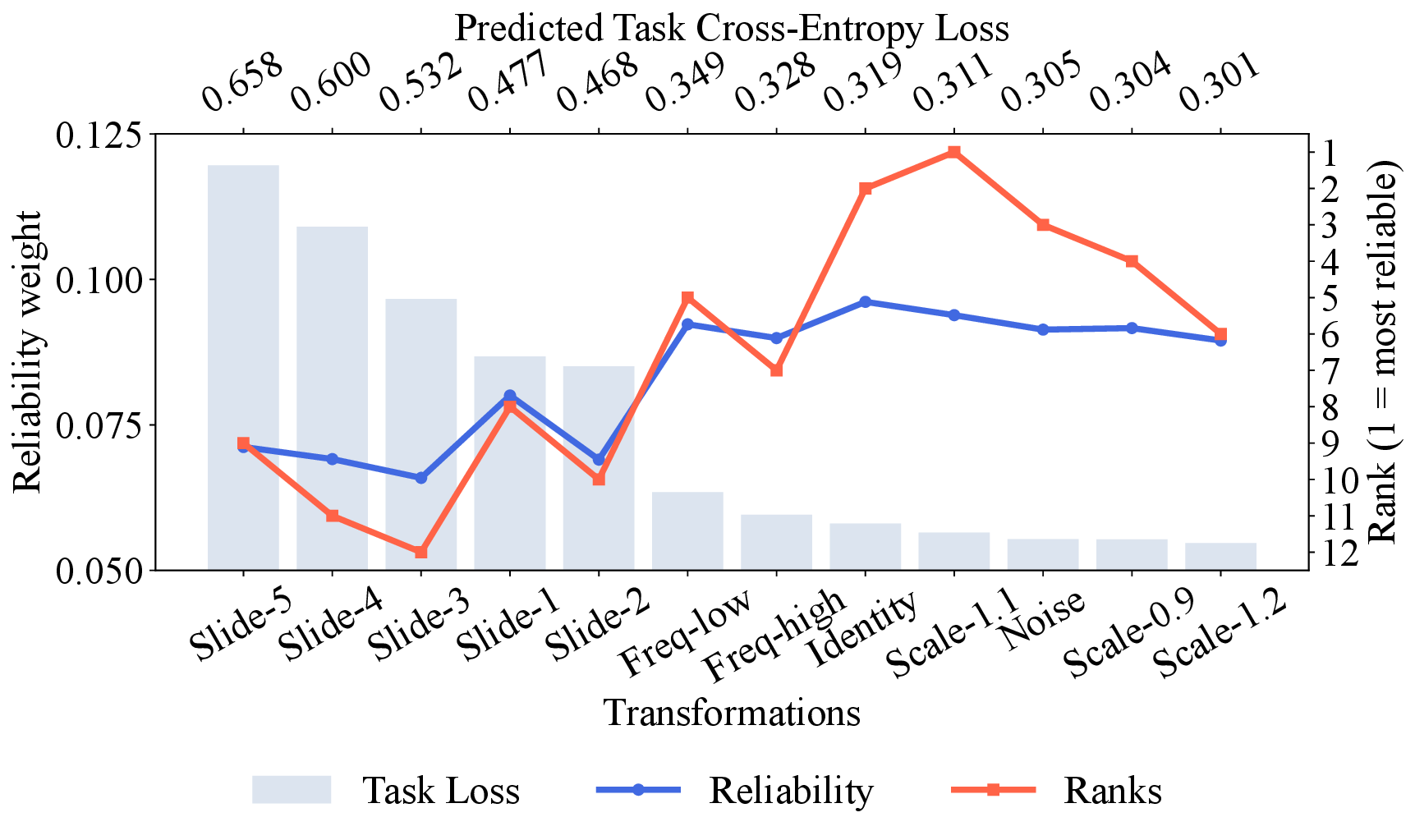} \label{fig:rankingclassification}}
\hfill
\subfigure[]{\includegraphics[width=.43\linewidth,clip]{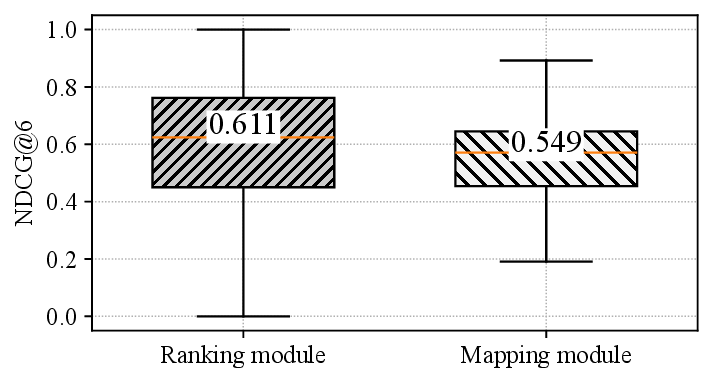} \label{fig:rankingregression}}
\caption{Evaluation of the learning-to-rank module for aggregation in classification and regression tasks. (a) Using subject S1 from the BNCI2014001 dataset as an example, three metrics were computed and averaged over all test trials for each of the twelve BFT-A transformations: (1) cross-entropy loss from the classifier $h(\cdot)$, (2) reliability weights from the ranking module $r(\cdot)$, and (3) integer-like ranks from the mapping module $m(\cdot)$. All weights were normalized to sum up to one; and (b) Using subject S1 from the Driving dataset as the test set as an example, we compared the statistics of NDCG@6, as the metric of the ranking performance of the reliability of the top-half of the transformations between: (1) ranking module's outputs, against the ground-truth task MSE loss; and (2) mapping module's outputs, against the ground-truth task MSE loss.}
\label{fig:ranking}
\end{figure*}

Additional ablation results on the learning-to-rank module for classification and regression tasks are presented in Fig.~\ref{fig:ranking}. Because the ranking module is trained only on source subjects, its benefit at test time relies on the branch-reliability structure remaining informative after a subject shift. The consistent gains of both BFT variants on held-out subjects (Tables~\ref{tab:Zhou2016} and~\ref{tab:BNCIandHG}) indicate that this structure does transfer, even though the base model itself is not adapted. Two conditions are therefore required for effective target-domain aggregation: a preserved reliability ranking, and adequate transformation coverage. When the deployed shift is not captured by the transformation bank, the expected benefit diminishes. Additional analysis can be found in Supplementary Material.

\begin{table}[!t]
\centering
\caption{Closeness between the synthetic pretraining scores and the real task losses, as means $\pm$ standard deviations across $n$ subject-seed units.}
\label{tab:loss_distribution}
\resizebox{\linewidth}{!}{%
\begin{tabular}{ll|cc|cc|c}
\toprule
\multirow{2}{*}{Dataset} & \multirow{2}{*}{Method} & \multicolumn{2}{c|}{Norm. spread} & \multirow{2}{*}{Hist. $L_1$} & \multirow{2}{*}{Wasserstein} & \multirow{2}{*}{$n$} \\
\cmidrule{3-4}
& & Real & Synthetic & & & \\
\midrule
Zhou2016 & BFT-A & 0.940$_{\pm0.080}$ & 0.968$_{\pm0.047}$ & 1.243$_{\pm0.144}$ & 0.156$_{\pm0.025}$ & 12 \\
Zhou2016 & BFT-D & 0.947$_{\pm0.067}$ & 0.967$_{\pm0.032}$ & 1.039$_{\pm0.221}$ & 0.114$_{\pm0.046}$ & 12 \\
BNCI2014001 & BFT-A & 1.000$_{\pm0.001}$ & 0.992$_{\pm0.017}$ & 1.120$_{\pm0.106}$ & 0.133$_{\pm0.022}$ & 27 \\
BNCI2014001 & BFT-D & 1.000$_{\pm0.001}$ & 0.956$_{\pm0.015}$ & 0.764$_{\pm0.113}$ & 0.063$_{\pm0.022}$ & 27 \\
Driving & BFT-A & 1.000$_{\pm0.000}$ & 0.939$_{\pm0.020}$ & 0.593$_{\pm0.073}$ & 0.045$_{\pm0.011}$ & 45 \\
Driving & BFT-D & 1.000$_{\pm0.000}$ & 0.986$_{\pm0.028}$ & 0.629$_{\pm0.209}$ & 0.045$_{\pm0.027}$ & 45 \\
\bottomrule
\end{tabular}}
\end{table}

\begin{table*}[t]
\centering
\caption{Quantization accuracy (FP32/INT8, on Zhou2016 S1 over three seeds) and per-sample TTA latency (FP32, after warm-up), on an Intel Xeon 8176 CPU and an NVIDIA RTX 3090 GPU. Each latency step runs on the listed device, except EA and transform generation, which run on the CPU. A dash denotes a step not applicable to the method.}
\label{tab:quantization}
\begin{tabular}{c|c|cc|c|cccccc|c}
\toprule
& & \multicolumn{2}{c|}{Accuracy (\%)} & & \multicolumn{7}{c}{Per-sample latency (ms)} \\
Approach & $K$ & FP32 & INT8 & Device & EA & Transform & Forward & Ranking & Aggreg. & Backprop. & Total \\
\midrule
EEGNet & 1 & 80.95$_{\pm2.77}$ & 80.95$_{\pm2.10}$ & -- & -- & -- & -- & -- & -- & -- & -- \\
\midrule
\multirow{2}{*}{T-TIME} & \multirow{2}{*}{1} & \multirow{2}{*}{83.75$_{\pm0.40}$} & \multirow{2}{*}{--} & CPU & 4.27 & -- & 1.30 & -- & -- & 13.76 & 19.33 \\
 & & & & GPU & 4.27 & -- & 0.48 & -- & -- & 4.41 & 9.16 \\
\midrule
\multirow{2}{*}{BFT-A} & \multirow{2}{*}{12} & \multirow{2}{*}{84.03$_{\pm2.74}$} & \multirow{2}{*}{83.19$_{\pm3.43}$} & CPU & 4.27 & 2.60 & 3.77 & 0.15 & 0.33 & -- & 11.12 \\
 & & & & GPU & 4.27 & 2.60 & 0.51 & 0.17 & 0.34 & -- & 7.89 \\
\midrule
\multirow{2}{*}{BFT-D} & \multirow{2}{*}{10} & \multirow{2}{*}{82.63$_{\pm2.21}$} & \multirow{2}{*}{82.07$_{\pm0.79}$} & CPU & 4.27 & 0.02 & 1.38 & 0.15 & 0.22 & -- & 6.04 \\
 & & & & GPU & 4.27 & 0.02 & 0.64 & 0.17 & 0.34 & -- & 5.44 \\
\bottomrule
\end{tabular}
\end{table*}

In Fig.~\ref{fig:rankingclassification}, transformations with lower task losses generally received higher reliability weights, although the magnitude differences were often subtle. The rank-based conversion amplified these distinctions, consistent with the mapping-ablation results in Table~\ref{tab:mappingablation}.

In Fig.~\ref{fig:rankingregression} for regression tasks, the ranking module achieved a median Normalized Discounted Cumulative Gain (NDCG) score of $0.611$ across test trials, considering the top half of the twelve transformations. Although the variation across trials was substantial, the performance remained substantially better than random ranking. Interestingly, we empirically observed that the ranking module's outputs slightly outperformed those of the mapping module.

The mapping module $m(\cdot)$ supplies a generic inductive bias from continuous reliability values to a differentiable ranking space. Synthetic pretraining establishes this rank geometry, while the subsequent source-domain training calibrates $r(\cdot)$ to the reliability of real EEG branches. At inference, aggregation uses the calibrated outputs of $r(\cdot)$, and $m(\cdot)$ is required only for rank-structured source training.

Because $m(\cdot)$ is pretrained on synthetic score vectors rather than real EEG, we verify that this surrogate occupies the same operating region as the real task losses it must later rank. It need not match any single dataset, since pretraining only teaches a generic map from continuous reliability values to a rank order. Within each subject-seed unit, we rescale the losses to $[0,1]$ and report in Table~\ref{tab:loss_distribution} the normalized spread of each distribution, so that close real and synthetic values indicate similar coverage, together with the histogram $L_1$ and Wasserstein distances between the real and synthetic distributions, where smaller values mean the two are more alike. The two spreads are nearly identical and the distances are small. The synthetic surrogate therefore covers the same operating region as the real losses, which is the property the mapping module needs.

\subsection{Quantization for Deployment}\label{sect:comptime}

In practice, neural network models for decoding in BCIs must operate under strict latency and memory constraints for edge computing~\cite{Syu2023, Chen2019}. Therefore, model quantization can reduce computational cost and storage while lowering inference latency~\cite{Schneider2020, Wang2020b}. We evaluated under reduced precision by applying post-training static quantization~\cite{Jacob2018}. Specifically, model weights trained on the source data were converted from 32-bit floating-point to 8-bit integer precision, using the training data. We tested the model on an NVIDIA GeForce RTX 3090 GPU and an Intel Xeon 8176 CPU.

Table~\ref{tab:quantization} jointly reports quantization accuracy on Zhou2016 S1 over three seeds and the per-sample TTA latency on both a CPU and a GPU. BFT-A and BFT-D retained most of their FP32 accuracy after INT8 conversion, whereas T-TIME updates model parameters during deployment and therefore does not follow a fixed post-training INT8 inference graph. Because the transformed branches are independent, BFT evaluates them in a single batched forward rather than sequentially, so the multiple views add little over one forward pass on the GPU and only a small multiple of it on the CPU. Both variants were therefore faster per sample than T-TIME, whose latency is dominated by the backpropagation-based parameter update. This advantage is most pronounced on the CPU, the more representative edge setting. BFT also remains compatible with post-training quantization. Preprocessing is excluded because it depends on the acquisition system.

These measurements establish a hardware-specific comparison rather than a universal latency claim: actual speed depends on batching, software kernels, and the target processor. On integer-optimized edge hardware, INT8 inference may provide additional gains~\cite{Schneider2020,Lai2018ARM}.

\section{Conclusions}\label{sect:conclusions}

This paper has proposed BFT, which reframes test-time adaptation for EEG decoding as a prediction-level operation rather than a parameter update. Reliability-ranked transformations of each trial are aggregated in a single forward pass to suppress inference uncertainty. BFT can deploy on quantized, resource-constrained edge devices where backpropagation-based TTA is impractical. More broadly, the reliability structure among transformed views, learned once on source data, is a transferable and underused signal for adaptation that never touches the model. To our knowledge, this is the first backpropagation-free TTA framework that unifies classification and regression for plug-and-play EEG-based BCIs.

The following directions can be considered in future research:
\begin{enumerate}
\item Label distribution shift: Addressing label distribution shift remains challenging without access to labeled target-domain data. Only a few approaches are applicable in this setting, and further investigation is needed.
\item Asynchronous BCIs: Adapting to asynchronous BCIs~\cite{Liu2026}, where the onset of trials are not explicitly marked, remains an open problem for TTA.
\item Trial rejection: Incorporating out-of-distribution detection~\cite{Liu2026} to identify/reject unreliable or corrupted test samples is a promising direction for improving robustness.
\end{enumerate}

\bibliographystyle{IEEEtran} \bibliography{bft}

\makeatletter
\let\maketitle\SAVEDmaketitle
\let\@maketitle\SAVEDatmaketitle
\makeatother
\setcounter{section}{0}
\setcounter{subsection}{0}
\setcounter{equation}{0}
\setcounter{figure}{0}
\setcounter{table}{0}
\renewcommand{\thesection}{S-\Roman{section}}
\renewcommand{\thesubsection}{\thesection-\Alph{subsection}}
\renewcommand{\theequation}{S\arabic{equation}}
\renewcommand{\thefigure}{S\arabic{figure}}
\renewcommand{\thetable}{S\arabic{table}}

\title{Supplementary Material for\\``Backpropagation-Free Test-Time Adaptation for Lightweight EEG-Based Brain-Computer Interfaces''}

\author{Siyang~Li, Jiayi~Ouyang, Zhenyao~Cui, Ziwei~Wang, Tianwang~Jia, Feng~Wan, and Dongrui~Wu, \IEEEmembership{Fellow,~IEEE}}

\markboth{IEEE JOURNAL OF BIOMEDICAL AND HEALTH INFORMATICS, 2026}
{LI \MakeLowercase{\textit{et al.}}: BACKPROPAGATION-FREE TEST-TIME ADAPTATION FOR LIGHTWEIGHT EEG-BASED BCIS (SUPPLEMENTARY MATERIAL)}
\maketitle

\section{Theoretical Foundation for BFT}
\label{sect:theory}

This section of the Supplementary Material gives a variance-based justification for BFT. It models transformations as draws from a design distribution and shows the conditions under which aggregating their predictions reduces variance. For deterministic BFT-D, the fixed mask bank is one realized design whose branch identities are retained at deployment, and no per-trial mask resampling is implied. Variance reduction yields more stable outputs but does not by itself guarantee higher accuracy under every domain shift.

\subsection{Label-Preserving Test-Time Randomization}
\label{sec:label_preserving}

\begin{definition}[Test-Time Randomization and Aggregation]
\label{def:sampling}
Let $\zeta$ index a label-preserving member of the transformation design distribution (e.g., a BFT-A transformation or a candidate structured feature mask). Fix an input $\mathbf x$. Define the scalar prediction under $\zeta$ as
\begin{align}
  f(\zeta;\mathbf{x})
  :=
  \mathbb{E} \left[y \middle| \mathbf x; \zeta \right]
  \in \mathbb R,
  \label{eq:def_f_omega}
\end{align}
where the expectation is taken with respect to the model-induced predictive distribution. In the notation of the main text, $f(\zeta_k;\mathbf x)$ corresponds to the prediction $h(\mathbf{z}^{(k)})$ of the $k$-th transformed input, and the weights $\mathbf w$ introduced below are those produced by the ranking module $r(\cdot)$.

To quantify prediction variability across the transformation design, single-branch uncertainty is measured via the variance:
\begin{align}
  V_0 := \mathrm{Var}_{\zeta} \big(f(\zeta;\mathbf{x})\big). 
  \label{eq:def_V0}
\end{align}

Let $\zeta_1,\dots,\zeta_K$ denote the selected design members and set $f_k:=f(\zeta_k;\mathbf{x})$. The $k$-th member defines the $k$-th test-time branch. For analysis, the branch design may be treated as independently and identically distributed (i.i.d.). More generally, we allow $\zeta_k\sim \mathcal A_k$ with branch-specific distributions $\{\mathcal A_k\}_{k=1}^K$. BFT-D fixes the resulting bank before ranking-module training and uses the same bank at test time.

A learning-to-rank module outputs weights $\mathbf w=(w_1,\dots,w_K)$ with $w_k\ge 0$ and $\sum_{k=1}^K w_k=1$, and the aggregation is
\begin{align}
  \hat f_{\mathbf{w}}(\mathbf{x})
  :=
  \sum_{k=1}^{K} w_k f_k.
  \label{eq:def_fhat_weighted}
\end{align}
The following deduction treats the realized weight vector $\mathbf{w}$ as deterministic. All variances are with respect to the transformation-design model $(\zeta_1,\dots,\zeta_K)$, not repeated random masks during deployed BFT-D inference.
\end{definition}

\subsection{Variance Decomposition for Weighted Aggregation}
\label{subsec:var_decomp}

\begin{lemma}[Exact Variance Decomposition]
\label{lem:var_exact}
Fix an input $\mathbf{x}$. Let $f_k:=f(\zeta_k;\mathbf{x})$ be square-integrable random variables induced by the joint test-time randomness, and define $\mu_k:=\mathbb{E}[f_k]$ for $k=1,\dots,K$. Let $\hat{f}_{\mathbf{w}}:=\sum_{k=1}^K w_k f_k$ with deterministic weights $\mathbf{w}$. Then we obtain
\begin{align}
  \mathrm{Var}\!\left(\hat f_{\mathbf w}\right)
  =
  \sum_{k=1}^{K} w_{k}^{2}\,\mathrm{Var}(f_k)
  +
  \sum_{i\ne j} w_i w_j\,\mathrm{Cov}(f_i,f_j).
  \label{eq:var_weighted_exact}
\end{align}
\end{lemma}

\noindent\textbf{Proof.}
\begin{align*}
  \mathrm{Var}(\hat f_{\mathbf {w}})
  &=
  \mathbb E\left[\Big(\hat f_{\mathbf{w}}-\mathbb{E}[\hat f_{\mathbf{w}}]\Big)^2\right]
  =
  \mathbb E\left[\Big(\sum_{k=1}^K w_k(f_k-\mu_k)\Big)^2\right] \\
  &=
  \sum_{i=1}^K\sum_{j=1}^K w_i w_j\,
  \mathbb E[(f_i-\mu_i)(f_j-\mu_j)]
  \\
  &=
  \sum_{i=1}^K\sum_{j=1}^K w_i w_j\,\mathrm{Cov}(f_i,f_j),
\end{align*}
which yields \eqref{eq:var_weighted_exact} by separating diagonal and off-diagonal terms. \hfill$\blacksquare$

\subsection{Homogeneous Variance Case}
\label{subsec:homo_var}

\begin{assumption}[Homogeneous Prediction Variance]
\label{ass:homo_var}
Assume
\begin{align}
  \mathrm{Var}(f_k)=\sigma^2,
  \qquad
  k=1,\dots,K.
  \label{eq:homo_var}
\end{align}
If $\sigma^2=0$, the prediction is deterministic and variance reduction is trivial. Otherwise, the correlations defined below are well-defined.
\end{assumption}

\begin{theorem}[Uncertainty Reduction under Homogeneous Variance]
\label{thm:strong_reduction}
Define $\rho_{ij}:=\mathrm{Corr}(f_i,f_j)$ for $i\ne j$.
Under Assumption~\ref{ass:homo_var},
\begin{align}
  \mathrm{Var} \left(\hat f_{\mathbf w}(\mathbf x)\right)
  =
  \sigma^2\sum_{k=1}^{K} w_k^2
  +
  \sigma^2\sum_{i\ne j} w_i w_j \rho_{ij}.
  \label{eq:var_homo_exact}
\end{align}
Let
\begin{align}
  \rho_{\max}
  :=
  \max_{\substack{i\neq j\\ i,j\in\{1,\dots,K\}}}
  |\rho_{ij}|
  \in[0,1].
  \label{eq:rhomax_def}
\end{align}
Then
\begin{align}
  \mathrm{Var}\!\left(\hat f_{\mathbf w}(\mathbf x)\right)
  \le
  \sigma^2\Big(
    \rho_{\max}
    +
    (1-\rho_{\max})\sum_{k=1}^{K} w_k^2
  \Big)
  \le
  \sigma^2.
  \label{eq:var_homo_bound}
\end{align}
Moreover, $\mathrm{Var}(\hat f_{\mathbf w}(\mathbf x))<\sigma^2$ whenever $\rho_{\max}<1$ and $\sum_{k=1}^K w_k^2<1$.
\end{theorem}

\noindent\textbf{Proof.}
By Lemma~\ref{lem:var_exact} and Assumption~\ref{ass:homo_var},
$\mathrm{Cov}(f_i,f_j)=\rho_{ij}\sigma^2$ for $i\ne j$, which gives
\eqref{eq:var_homo_exact}. For the bound, use $|\rho_{ij}|\le \rho_{\max}$ and $\sum_{i\ne j} w_i w_j = 1-\sum_k w_k^2$:
\begin{align*}
\sum_{i\ne j} w_i w_j \rho_{ij}
\le \sum_{i\ne j} w_i w_j |\rho_{ij}|
\le \rho_{\max}\Big(1-\sum_{k=1}^K w_k^2\Big),
\end{align*}
where we used $(\sum_k w_k)^2=1=\sum_k w_k^2+\sum_{i\ne j} w_i w_j$. Substituting into \eqref{eq:var_homo_exact} yields \eqref{eq:var_homo_bound}. \hfill$\blacksquare$

\begin{figure}[htpb] \centering
\includegraphics[width=1.02\linewidth,clip]{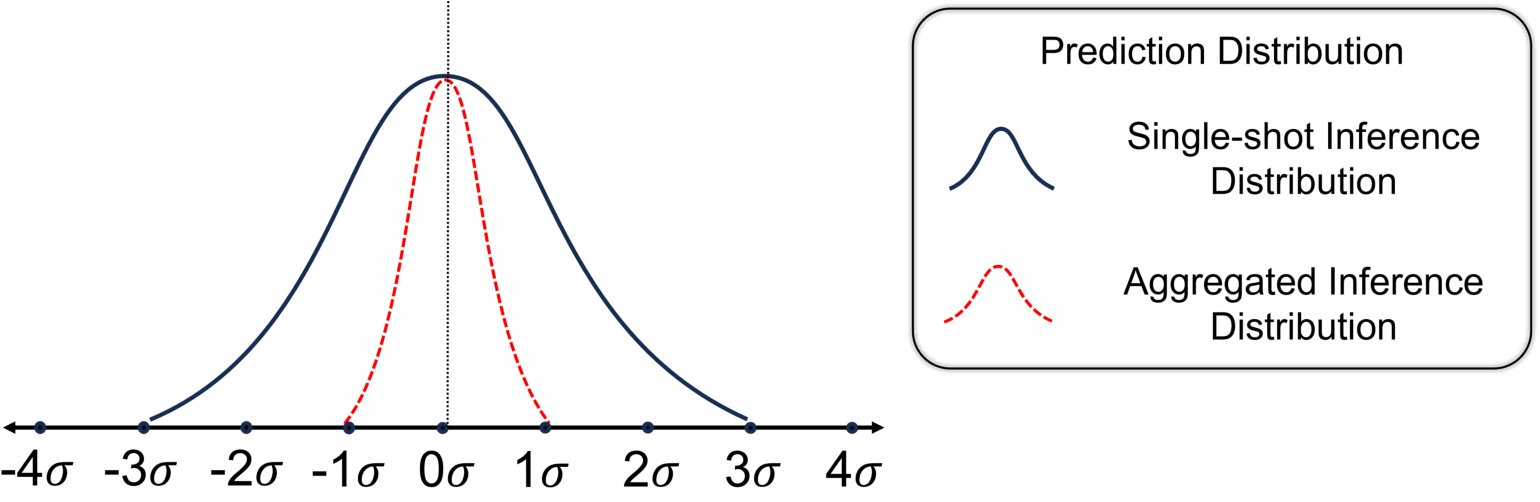}
\caption{Illustration of uncertainty reduction achieved through test-time transformations under the homogeneous variance assumption.} \label{fig:uncertainty}
\end{figure}

An illustrative visualization of Theorem~\ref{thm:strong_reduction} is provided in Fig.~\ref{fig:uncertainty}.

\subsection{Heterogeneous Variance Case}
\label{subsec:hetero_var}

In the heterogeneous-variance setting, the ensemble uncertainty is mainly affected by three factors: the worst-branch noise $\kappa$, the similarity between branches $\rho_{\max}$, and how spread the weights are $K_{\mathrm{eff}}$.

\begin{assumption}[Heterogeneous Prediction Variance]
\label{ass:finite_variance_controlled}
Fix an input $\mathbf x$. In the $k$-th branch (defined in Definition~\ref{def:sampling}), the prediction $f_k$ has heterogeneous variance:
\begin{align}
  \mathrm{Var}(f_k)=\sigma_k^2,
  \qquad
  0<\sigma_k^2\le \sigma_{\max}^2<\infty,
  \quad k=1,\dots,K.
  \label{eq:hetero_var}
\end{align}
Let $V_0:=\mathrm{Var}_{\zeta}\!\big(f(\zeta;\mathbf x)\big)$ be the single-shot test-time variance defined in~\eqref{eq:def_V0}, where $\zeta$ follows the randomization used in single-shot inference. Assume that the worst branch variance is controlled relative to $V_0$:
\begin{align}
  \sigma_{\max}^2 \le \kappa V_0,
  \qquad
  1 \le \kappa.
  \label{eq:var_spread_kappa}
\end{align}
\end{assumption}

\begin{theorem}[Uncertainty Reduction under Heterogeneous Variance]
\label{thm:unbiased_reduction}
Let $\rho_{\max}$ be defined in \eqref{eq:rhomax_def}. Under
Assumption~\ref{ass:finite_variance_controlled}, for any probability weights $\mathbf w$ (i.e., $w_k\ge 0$ and $\sum_{k=1}^K w_k=1$),
\begin{align}
  \mathrm{Var}\!\left(\hat f_{\mathbf w}(\mathbf x)\right)
  \;&\le\;
  \sigma_{\max}^2
  \Big(
    \rho_{\max}
    +
    (1-\rho_{\max})\!\sum_{k=1}^{K} w_k^2
  \Big)
  \nonumber\\
  &\le\;
  \kappa V_0
  \Big(
    \rho_{\max}
    +
    (1-\rho_{\max})\!\sum_{k=1}^{K} w_k^2
  \Big).
  \label{eq:thm2_weighted_bound}
\end{align}
Define the effective number of branches
\begin{align}
  K_{\mathrm{eff}}
  :=
  \frac{1}{\sum_{k=1}^K w_k^2}
  \in [1,K].
  \label{eq:Keff_def}
\end{align}
If $\rho_{\max}<1/\kappa$, then a sufficient condition for
$\mathrm{Var}\!\left(\hat f_{\mathbf w}(\mathbf x)\right) < V_0$ is
\begin{align}
  K_{\mathrm{eff}}
  \;>\;
  \frac{\kappa(1-\rho_{\max})}{1-\kappa\rho_{\max}}.
  \label{eq:Keff_condition_controlled}
\end{align}
\end{theorem}

\noindent\textbf{Proof.}
For $i\ne j$, let $\rho_{ij}:=\mathrm{Corr}(f_i,f_j)$. Then
\begin{align}
  |\mathrm{Cov}(f_i,f_j)|
  &=
  |\rho_{ij}|\sqrt{\mathrm{Var}(f_i)\mathrm{Var}(f_j)}
  \nonumber\\
  &\le
  \rho_{\max}\sqrt{\sigma_i^2\sigma_j^2}
  \le
  \rho_{\max}\sigma_{\max}^2.
  \label{eq:cov_bound}
\end{align}
Since $w_i w_j\ge 0$,
\begin{align}
  \sum_{i\ne j} w_i w_j\,\mathrm{Cov}(f_i,f_j)
  &\le
  \rho_{\max}\sigma_{\max}^2
  \sum_{i\ne j} w_i w_j
  \nonumber\\
  &=
  \rho_{\max}\sigma_{\max}^2
  \Big(1-\sum_{k=1}^K w_k^2\Big),
  \label{eq:offdiag_bound}
\end{align}
where we used $(\sum_k w_k)^2 = 1 = \sum_k w_k^2 + \sum_{i\ne j} w_i w_j$.
Moreover,
\begin{align}
  \sum_{k=1}^K w_k^2\,\mathrm{Var}(f_k)
  \;=\;
  \sum_{k=1}^K w_k^2\sigma_k^2
  \;\le\;
  \sigma_{\max}^2\sum_{k=1}^K w_k^2.
  \label{eq:diag_bound}
\end{align}
Combining \eqref{eq:offdiag_bound} and \eqref{eq:diag_bound} with \eqref{eq:var_weighted_exact} yields the first inequality in \eqref{eq:thm2_weighted_bound}. The second inequality follows from \eqref{eq:var_spread_kappa}.

For the sufficient condition, it is enough to ensure that the upper bound in \eqref{eq:thm2_weighted_bound} is strictly smaller than $V_0$, namely
\begin{align}
  \kappa\Big(
    \rho_{\max} + (1-\rho_{\max})\sum_{k=1}^{K} w_k^2
  \Big) < 1.
\end{align}
This inequality requires $1-\kappa\rho_{\max}>0$, i.e.,
$\rho_{\max}<1/\kappa$, and under this condition it is equivalent to
\begin{align}
  \sum_{k=1}^{K} w_k^2
  \;<\;
  \frac{1-\kappa\rho_{\max}}{\kappa(1-\rho_{\max})}.
\end{align}
Using $K_{\mathrm{eff}} = 1/\sum_{k=1}^K w_k^2$, we obtain
\eqref{eq:Keff_condition_controlled}. \hfill$\blacksquare$

The bound improves when branches are less correlated (small $\rho_{\max}$) and the weights are not overly concentrated (large $K_{\mathrm{eff}}$). These are sufficient conditions, not properties guaranteed by the architecture. They also identify failure modes: highly correlated branches, an unreliable source-trained ranking, or transformations that do not preserve the target label can remove the expected benefit.

\section{Robustness of the Reliability-Aware Aggregation}
\label{app:robustness}

Two design choices determine whether BFT helps, namely the diversity of the transformation bank and the quality of the learned ranking. This section reports two analyses that probe each choice.

\subsection{Transformation Coverage}
\label{app:coverage}

\begin{figure*}[htpb] \centering
\includegraphics[width=\linewidth,clip]{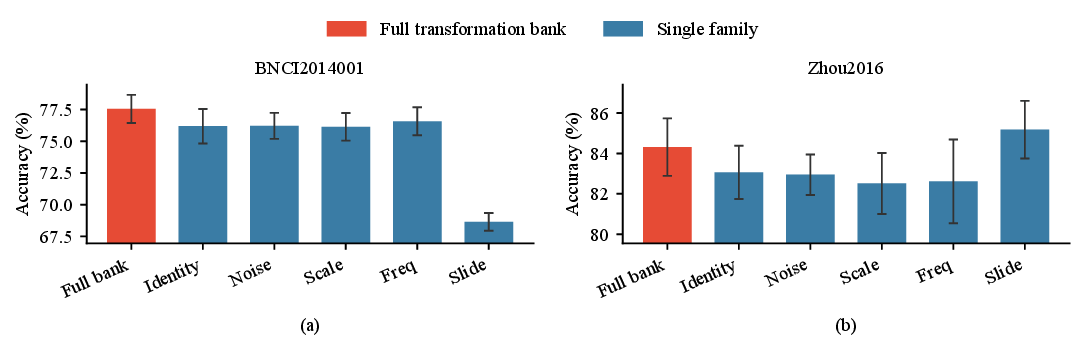}
\caption{Classification accuracy of the full BFT-A transformation bank compared with each single transformation family, on (a) BNCI2014001 and (b) Zhou2016. No single family is best on both datasets, so the full bank is the robust default.} \label{fig:transformation_family}
\end{figure*}

\begin{figure*}[htpb] \centering
\includegraphics[width=\linewidth,clip]{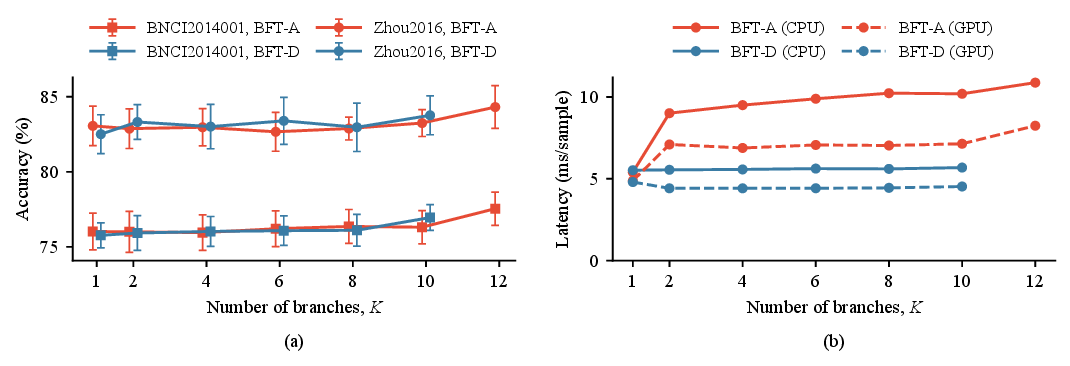}
\caption{Performance w.r.t. the number of branches $K$ in BFT, with EEGNet. Error bars in (a) are standard deviations over the seed-subject units. BFT-D is swept up to its operating point $K=10$, and BFT-A up to $K=12$. (a) Accuracy; and (b) Latency.} \label{fig:k_accuracy_latency}
\end{figure*}

BFT aggregates branches produced by a bank of transformation families, so a natural concern is whether the method depends on any single family. Fig.~\ref{fig:transformation_family} compares the full BFT-A bank against each single family on BNCI2014001 and Zhou2016. On BNCI2014001 the full bank reaches 77.55\%, whereas the sliding-window family alone collapses to 68.65\%. On Zhou2016 the sliding-window family alone is the strongest single choice at 85.18\%, yet it is the weakest on BNCI2014001, and no single family is best on both datasets. The full bank is therefore the robust default, because it does not rely on any one family suiting the deployed data.

Varying the number of branches $K$ showed that accuracy is not monotonic in $K$, as additional branches help only when they add useful views, as evident in Fig.~\ref{fig:k_accuracy_latency}. The operating points $K=12$ for BFT-A and $K=10$ for BFT-D gave the strongest measured accuracy for each variant under this protocol.

\subsection{Ranking Robustness}
\label{app:ranking}

\begin{table}[htpb]
\centering
\caption{Effect of degrading the learned ranking at test time on classification accuracy (\%), replacing it by Gaussian-perturbed, random, reversed, or shuffled rankings. Means $\pm$ standard deviations over the seed-subject units (27 for BNCI2014001, 12 for Zhou2016). The learned ranking is best in every case.}
\label{tab:ranking_robustness}
\resizebox{\linewidth}{!}{%
\begin{tabular}{l|cc|cc}
\toprule
\multirow{2}{*}{Ranking} & \multicolumn{2}{c|}{BNCI2014001} & \multicolumn{2}{c}{Zhou2016} \\
\cmidrule{2-5}
& BFT-A & BFT-D & BFT-A & BFT-D \\
\midrule
Learned & \textbf{77.52}$_{\pm1.09}$ & \textbf{77.01}$_{\pm0.83}$ & \textbf{84.31}$_{\pm1.43}$ & \textbf{83.76}$_{\pm1.29}$ \\
Gaussian noise & 75.75$_{\pm0.96}$ & 75.87$_{\pm1.23}$ & 83.41$_{\pm1.17}$ & 82.52$_{\pm1.30}$ \\
Random & 75.85$_{\pm1.36}$ & 76.16$_{\pm1.04}$ & 83.54$_{\pm1.41}$ & 82.73$_{\pm1.07}$ \\
Reverse & 75.46$_{\pm1.09}$ & 75.75$_{\pm0.96}$ & 83.01$_{\pm1.23}$ & 82.56$_{\pm1.18}$ \\
Shuffle & 76.24$_{\pm1.12}$ & 76.18$_{\pm1.18}$ & 83.73$_{\pm1.22}$ & 82.96$_{\pm1.44}$ \\
\bottomrule
\end{tabular}}
\end{table}

The reliability ranking is trained only on source subjects, so it is important to confirm that the ranking itself, and not the averaging alone, drives the improvement at test time. We replace the learned ranking with four degraded rankings, namely Gaussian-perturbed, random, reversed, and shuffled, and measure the resulting accuracy. Table~\ref{tab:ranking_robustness} shows that the learned ranking is best in every dataset and variant, and that every degraded ranking lowers accuracy, by up to 2.06\% on BNCI2014001 for BFT-A when the ranking is reversed. The reliability structure learned on the source subjects therefore remains informative after a subject shift, and the ranking, not the uniform average, is responsible for the gain.

\subsection{Regression Aggregation Rule}
\label{app:regr_agg}

\begin{table}[htpb]
\centering
\caption{Regression aggregation rules on the Driving dataset, as the change in RMSE and CC relative to the top-half mean used in the main text. Every alternative stays within 0.005 RMSE and 0.003 CC of it, far below the across-seed standard deviation. Averaged over three seeds and all subjects.}
\label{tab:regression_aggregation}
\scalebox{0.87}{\begin{tabular}{l|cc|cc}
\toprule
\multirow{2}{*}{Aggregation rule} & \multicolumn{2}{c|}{BFT-A} & \multicolumn{2}{c}{BFT-D} \\
\cmidrule{2-5}
& $\Delta$RMSE & $\Delta$CC & $\Delta$RMSE & $\Delta$CC \\
\midrule
Top-half mean (used) & 0.0000 & 0.0000 & 0.0000 & 0.0000 \\
Reliability-weighted, all branches & $-$0.0002 & $+$0.0030 & $+$0.0019 & $+$0.0013 \\
Reliability-weighted, top half & 0.0000 & 0.0000 & 0.0000 & 0.0000 \\
Median & $-$0.0005 & $+$0.0012 & $+$0.0046 & $+$0.0013 \\
Trimmed mean & $-$0.0003 & $+$0.0024 & $+$0.0046 & $+$0.0015 \\
\bottomrule
\end{tabular}}
\end{table}

For regression, BFT averages the top-ranked half of the transformed predictions (Section~III-D of the main text). We tested whether this discards useful information carried by the continuous reliability scores. On the Driving dataset we therefore compared the top-half mean against alternatives that do use the continuous scores or a different pooling rule, namely a reliability-weighted mean over all branches, a reliability-weighted mean over the top half, the median, and a trimmed mean. Table~\ref{tab:regression_aggregation} reports each rule as the change in RMSE and CC relative to the top-half mean, so a value near zero means the rule matches it. Every alternative, including the continuous reliability weighting, stays within 0.005 in RMSE and 0.003 in CC of the top-half mean for both BFT-A and BFT-D. These differences are far smaller than the across-seed standard deviation, which is about 0.010 in RMSE and 0.017 in CC, so no rule is reliably better. This matches the design: the ranking objective supervises only the order of the branches, not calibrated inverse-error magnitudes, so the continuous scores carry no additional accuracy for continuous outputs, and the tuning-free top-half mean is preferred.

\end{document}